\documentclass[aps,prd,preprint,nofootinbib,11pt]{revtex4}
\usepackage{amsfonts}
\usepackage{mathrsfs}
\usepackage{graphicx}
\usepackage{amsmath}
\usepackage{amssymb}
\usepackage{subfigure}
\usepackage{epsfig}
\usepackage{graphicx}
\usepackage{color}
\newcommand{\bqa}{\begin{eqnarray}}
\newcommand{\eqa}{\end{eqnarray}}
\newcommand{\beq}{\begin{equation}}
\newcommand{\eeq}{\end{equation}}
\allowdisplaybreaks[1]
\graphicspath{{fig/}{dia/}} \DeclareGraphicsExtensions{.eps}

\begin{document}
\title{The tetra-heavy baryonium spectra}

\author{Bing-Dong Wan$^{1,2}$\footnote{wanbingdong16@mails.ucas.ac.cn} and Cong-Feng Qiao$^{2}$\footnote{qiaocf@ucas.ac.cn, corresponding author}\vspace{+3pt}}

\affiliation{$^1$  School of Fundamental Physics and Mathematical Sciences, Hangzhou Institute for Advanced Study, UCAS, Hangzhou 310024, China\\
$^2$ School of Physics, University of Chinese Academy of Science, Yuquan Road 19A, Beijing 100049, China}

\author{~\\~\\}

\begin{abstract}
\vspace{0.3cm}
In this work, we investigate the spectra of the prospective tetracharm and tetrabottom baryonium, viz. the baryon-antibaryon states, in the framework of QCD sum rules. The non-perturbative contributions up to dimension 10 are taken into account. Numerical results indicate that there might exist 4 possible tetracharm baryonium states with masses $(7.33\pm0.12)$, $(7.42\pm0.13)$, $(7.68\pm0.17)$, and $(7.76\pm0.12)$ GeV for the quantum numbers of $0^{-+}$, $1^{--}$, $0^{++}$, and $1^{++}$, respectively. The corresponding tetrabottom partners are found lying respectively at $(19.41\pm 0.15)$, $(19.48\pm 0.15)$, $(19.77\pm 0.19)$ and $(19.84\pm 0.19)$ GeV.
Additionally, we also analyze the spectra of the $\bar{\Xi}_{bc}\Xi_{bc}$ and $\bar{\Xi}_{cc}\Xi_{bb}$ baryonium states, which lie in the range of $13.77-14.40$ GeV and $13.39-14.14$ GeV, respectively. The possible baryonium decay modes are analyzed, which are hopefully measurable in LHC experiments.
\end{abstract}
 \maketitle
\newpage

\section{Introduction}

In the framework of quark model~\cite{GellMann:1964nj,Zweig}, the discovery of exotic states, i.e. the hadronic structures beyond the conventional meson ($\bar{q}q$) and baryon ($qqq$), greatly enriches the hadron family and our knowledge of the nature of QCD. So far, more than thirty exotic states or candidates, such as X(3872)~\cite{Choi:2003ue}, have been observed in experiments, and new ones tend to appear more frequently. Needless to say, to understand the properties of the exotic states and find more possible exotic states is one of the most intriguing research topics in hadron physics, which attracts more and more interests from theorists and experimentalists.

In 2020, using proton-proton collision data at centre-of-mass energies of $\sqrt{s}=7$, 8 and 13 TeV, LHCb Collaboration observed a narrow structure in $J/\psi$-pair invariant mass of about $6.9$ GeV with the significance greater than $5\; \sigma$~\cite{Aaij:2020fnh}. Furthermore, a broad structure just above double $J/\psi$ threshold and a vague structure around $7.2$ GeV were also reported. This is the first clear structures that have been observed in the $J/\psi$-pair mass spectrum, and has immediately inspired extensive discussions on the novel tetracharm hadronic structure. Very recently, the $X(6900)$ was confirmed by ATLAS and CMS Collaborations, and two new structures, the $X(6600)$ and $X(7200)$ have been observed in $J/\psi$-pair invariant mass spectrum~\cite{ATLAS,CMS}. If those structures are further confirmed to be hadronic structures, the tetracharm states rather merely some kinematic effect, the new finding will be considered as a huge breakthrough in the exploration of hadron spectroscopy.

In the literature, some studies on the tetracharm (bottom) states have been conducted~\cite{Iwasaki:1976cn,Chao:1980dv,Ader:1981db,Karliner:2016zzc,Barnea:2006sd,Debastiani:2017msn,Liu:2019zuc,Chen:2016jxd,Wang:2019rdo,Lloyd:2003yc,Anwar:2017toa,Wang:2017jtz,Jin:2020jfc,Lu:2020cns,Yang:2020rih,Wang:2020ols,Albuquerque:2020hio,Sonnenschein:2020nwn,Giron:2020wpx,liu:2020eha,Gong:2020bmg,Weng:2020jao,Dong:2020nwy,Yang:2020atz,Wang:2020wrp,Wan:2020fsk,Yang:2020wkh,Liang:2021fzr,Zhao:2020zjh,Mutuk:2022nkw,Wang:2022yes,An:2022qpt,Lu:2022myk,Wu:2022qwd,Wang:2021mma,Zhuang:2021pci}, and most of these theoretical studies fall in the experimental measurement about the $X(6900)$ and the broad structure.
However, the structure $X(7200)$ whose mass measured to be $7.29$ GeV is higher than double $J/\psi$ threshold by $1.1$ GeV or so, much larger than the typical energy gap between ground and excited states.
On the other hand, $X(7200)$ lies very close to the $\bar{\Xi}_{cc}\Xi_{cc}$ threshold $7.24$ GeV \cite{LHCb:2017iph}, and it can be naturally explained as $\bar{\Xi}_{cc}\Xi_{cc}$ hexaquark molecular state.
In Ref~\cite{Liu:2020tqy}, it has been discussed that the two doubly charmed baryon channel could also affect the mass spectra of fully charmed tetraquark states, and the pseudoscalar baryon-antibaryon molecular state is investigated in Ref.~\cite{Wang:2021wjd}, and the predicted mass is around $7.2$ GeV.

Actually, the story of the investigation of baryon-antibaryon states can date back to the 1940s, when Fermi and Yang proposed that $\pi$-mesons may be a composite particle of nucleon-antinucleon \cite{Fermi:1949voc}, and their scenario was later on replaced by the quark model. Entering the 21st century, the heavy baryon-antibaryon hadronic structures, and hence the term of baryonium, were proposed and employed to explain the extraordinary nature of $Y(4260)$ \cite{Qiao:2005av, Qiao:2007ce} and other charmonium-like states observed in experiments. Later on, more investigations on baryonium are performed from various aspects \cite{Chen:2011cta, Chen:2013sba,Wan:2019ake,Wan:2021vny,Chen:2016ymy,Liu:2007tj,Wang:2021qmn,Wang:2021pua}.

In this work, we evaluate the tetracharm and tetrabottom baryonioum by using the the model with the independent Shifman, Vainshtein and Zakharov (SVZ) sum rule technique \cite{Shifman}, and the $\bar{\Xi}_{bc}\Xi_{bc}$ and $\bar{\Xi}_{cc}\Xi_{bb}$ baryonium states are also investigated. The SVZ sum rule, viz the QCD sum rule (QCDSR), has some peculiar advantages in exploring hadron properties involving nonperturbative QCD. Rather than a phenomenological model, QCDSR is a QCD based theoretical framework which incorporates nonperturbative effects universally order by order and has already achieved a lot in the study of hadron spectroscopy and decays. To establish the sum rule, the starting point is to construct the proper interpolating currents corresponding to the hadron of interest, which possesses the foremost information about the concerned hadron, like quantum numbers and the constituent quark or gluon. By using the currents, one can then construct the two-point correlation function, which has two representations: the QCD representation and the phenomenological representation. Equating these two representations, the QCD sum rules will be formally established, from which the hadron mass and decay width may be deduced.

The rest of the paper is arranged as follows. After the introduction, a brief interpretation of QCD sum rules and some primary formulas in our calculation are presented in Sec. \ref{Formalism}. The numerical analysis and results are given in Sec. \ref{Numerical}. In Sec. \ref{Decay}, possible decay modes of tetra-heavy baryonium states are investigated. The last part is left for conclusions and discussions.

\section{Formalism}\label{Formalism}

To evaluate the mass spectrum of tetracharm/tetrabottom baryonium in QCDSR, the appropriate currents coupling to the states have to be constructed. The lowest order interpolating currents for tetracharm/tetrabottom baryonium can be respectively constructed as
\begin{eqnarray}\label{current_lambda}
j(x)&=& \bar{\eta}(x) \Gamma \eta(x) \;. \label{Ja}
\end{eqnarray}
Here, we use the notion $\eta$ to represent the Dirac baryon fields  without free Lorentz indices, and $\Gamma=i\gamma_5$, $\gamma_\mu$, $I$, and $i\gamma_\mu\gamma_5$ for quantum numbers of $J^{PC}=0^{-+}$, $1^{--}$, $0^{++}$, and $1^{++}$, respectively. As shown in Ref. \cite{Chung:1981wm}, $\eta$ may takes the following quark structure :
\begin{eqnarray}\label{current_lambda_b}
\eta(x)&=&i \epsilon_{a b c}[ Q_a^{T}(x) C \gamma_5 q_b(x) ]Q_c(x) \; ,\label{Jabaryon}
\end{eqnarray}
where the subscripts $a$, $b$, and $c$ are color indices, $Q$ represents the heavy-quark $c$ or $b$, $q$ stands for the light-quark $u$ or $d$, and $C$ is the charge conjugation matrix.
In principle, there may be other currents couple to the tetracharm/tetrabottom baryonium, while the Eq.~(\ref{Ja}) is the simplest one. It should be note that exchanging $Q$ and $q$ in Eq.~(\ref{Jabaryon}) will break Fermi-Dirac statistics.

With the currents (\ref{Ja}), the two-point correlation function can be readily established, i.e.,
\begin{eqnarray}
\Pi(q^2) &=& i \int d^4 x e^{i q \cdot x} \langle 0 | T \{ j (x),\;  j^\dagger (0) \} |0 \rangle\;,\\
\Pi_{\mu\nu}(q^2) &=& i \int d^4 x e^{i q \cdot x} \langle 0 | T \{ j_\mu (x),\;  j_\nu^\dagger (0) \} |0 \rangle \;,
\end{eqnarray}
where, $j(x)$ and $j_\mu(x)$ are the relevant hadronic currents with $J = 0$ and 1, respectively, and $|0\rangle$ denotes the physical vacuum. For $j_\mu(x)$, the correlation function has the following Lorentz covariance form:
\begin{eqnarray}
\Pi_{\mu\nu}(q^2) &=&-\Big( g_{\mu \nu} - \frac{q_\mu q_\nu}{q^2}\Big) \Pi_1(q^2)+ \frac{q_\mu q_\nu}{q^2}\Pi_0(q^2)\;,
\end{eqnarray}
where the subscripts $1$ and $0$, respectively, denote the quantum numbers of the spin 1 and 0 mesons.

In the hadronic representation, after isolating the ground state contribution from the hexaquark state, the correlation function $\Pi^{X}_{J^{PC}}(q^2)$ can be expressed as a dispersion integral over the physical regime, i.e.,
 \begin{eqnarray}
\Pi^{X,\;phen}_{J^{PC}}(q^2) & = & \frac{(\lambda^X_{J^{PC}})^2}{(m^X_{J^{PC}})^2 - q^2} + \frac{1}{\pi} \int_{s_0}^\infty d s \frac{\rho^X_{J^{PC}}(s)}{s - q^2} \; , \label{hadron}
\end{eqnarray}
where the superscript $X$ denotes the lowest lying hexaquark state, $m^X_{J^{PC}}$ is the mass of $X$ with the quantum number of $J^{PC}$, $\rho^X_{J^{PC}}(s)$ is the spectral density that contains the contributions from higher excited states and the continuum states above the threshold $s_0$, and $\lambda^X$ is the coupling constant.

In the QCD representation, the dispersion relation can express the correlation function $\Pi^{X}_{J^{PC}}(q^2)$ as
 \begin{eqnarray}
\Pi^{X,\;OPE}_{J^{PC}} (q^2) = \int_{16 m_Q^2}^{\infty} d s
\frac{\rho^{X,\;OPE}_{J^{PC}}(s)}{s - q^2}\; ,
\label{OPE-hadron}
\end{eqnarray}
where $\rho^{X,\;OPE}_{J^{PC}}(s) = \text{Im} [\Pi^{X,\;OPE}_{J^{PC}}(s)] / \pi$, and
\begin{eqnarray}
\rho^{OPE}(s) & = & \rho^{pert}(s) + \rho^{\langle \bar{q} q
\rangle}(s) +\rho^{\langle G^2 \rangle}(s) + \rho^{\langle \bar{q} G q \rangle}(s)
+ \rho^{\langle \bar{q} q \rangle^2}(s) \nonumber\\
&+& \rho^{\langle G^3 \rangle}(s)
+ \rho^{\langle \bar{q} q \rangle\langle \bar{q} G q \rangle}(s)+ \rho^{\langle G^2 \rangle\langle \bar{q} q \rangle^2}(s)  . \label{rho-OPE}
\end{eqnarray}

\begin{figure}
\includegraphics[width=6.8cm]{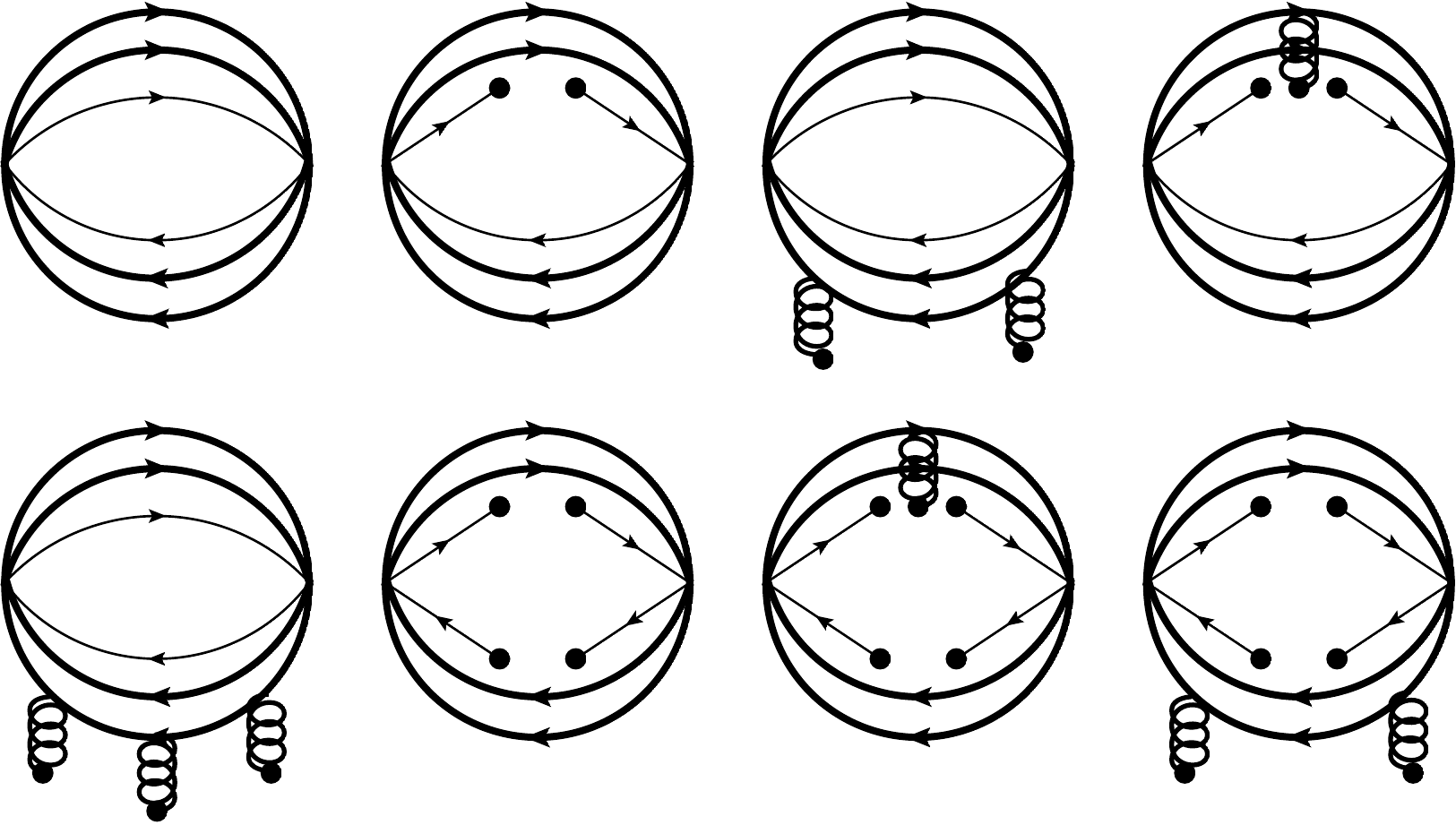}
\caption{The typical Feynman diagrams related to the correlation function, where the thick solid line stands for the heavy quark, the thin solid line represents the light quark, and the sprial line denotes the gluon.} \label{feyndiag}
\end{figure}

 To calculate the spectral density of the operator
product expansion (OPE) side, Eq. (\ref{rho-OPE}), the light quark and heavy-quark full propagators $S_{ij}^q(x)$ and $S_{ij}^Q(p)$ are employed, say
\begin{eqnarray}
S^q_{i j}(x) \! \! & = & \! \! \frac{i \delta_{i j} x\!\!\!\slash}{2 \pi^2
x^4} - \frac{\delta_{i j} m_q}{4 \pi^2 x^2} - \frac{i t^a_{i j} G^a_{\alpha\beta}}{32 \pi^2 x^2}(\sigma^{\alpha \beta} x\!\!\!\slash
+ x\!\!\!\slash \sigma^{\alpha \beta}) - \frac{\delta_{i j}}{12} \langle\bar{q} q \rangle + \frac{i\delta_{i j}
x\!\!\!\slash}{48} m_q \langle \bar{q}q \rangle - \frac{\delta_{i j} x^2}{192} \langle g_s \bar{q} \sigma \cdot G q \rangle \nonumber \\ &+& \frac{i \delta_{i j} x^2 x\!\!\!\slash}{1152} m_q \langle g_s \bar{q} \sigma \cdot G q \rangle - \frac{t^a_{i j} \sigma_{\alpha \beta}}{192}
\langle g_s \bar{q} \sigma \cdot G q \rangle
+ \frac{i t^a_{i j}}{768} (\sigma_{\alpha \beta} x \!\!\!\slash + x\!\!\!\slash \sigma_{\alpha \beta}) m_q \langle
g_s \bar{q} \sigma \cdot G q \rangle \;,\\
S^Q_{i j}(p) \! \! & = & \! \! \frac{i \delta_{i j}(p\!\!\!\slash + m_Q)}{p^2 - m_Q^2} - \frac{i}{4} \frac{t^a_{i j} G^a_{\alpha\beta} }{(p^2 - m_Q^2)^2} [\sigma^{\alpha \beta}
(p\!\!\!\slash + m_Q)
+ (p\!\!\!\slash + m_Q) \sigma^{\alpha \beta}] \nonumber \\
&+& \frac{i\delta_{i j}m_Q  \langle g_s^2 G^2\rangle}{12(p^2 - m_Q^2)^3}\bigg[ 1 + \frac{m_Q (p\!\!\!\slash + m_Q)}{p^2 - m_Q^2} \bigg] \nonumber \\ &+& \frac{i \delta_{i j}}{48} \bigg\{ \frac{(p\!\!\!\slash +
m_Q) [p\!\!\!\slash (p^2 - 3 m_Q^2) + 2 m_Q (2 p^2 - m_Q^2)] }{(p^2 - m_Q^2)^6}
\times (p\!\!\!\slash + m_Q)\bigg\} \langle g_s^3 G^3 \rangle \; ,
\end{eqnarray}
where, the vacuum condensates are clearly displayed. For more explanation on above propagator, readers may refer to Refs.~\cite{Wang:2013vex, Albuquerque:2013ija}. The Feynman diagrams corresponding to each term of Eq. (\ref{hadron}) are schematically shown in Fig.~\ref{feyndiag}.

Performing Borel transform on both Eq.~(\ref{hadron}) and Eq.~(\ref{OPE-hadron}), and matching the QCD side with the hadronic side of the correlation function $\Pi(q^2)$, one can finally obtain the mass of the hexaquark state,
\begin{eqnarray}
m^X_{J^{PC}}(s_0, M_B^2) = \sqrt{- \frac{L^{X}_{J^{PC},\;1}(s_0, M_B^2)}{L^{X}_{J^{PC},\;0}(s_0, M_B^2)}} \; . \label{mass-Eq}
\end{eqnarray}
Here $L_0$ and $L_1$ are respectively defined as
\begin{eqnarray}
L^{X}_{J^{PC},\;0}(s_0, M_B^2) =  \int_{16m_Q^2}^{s_0} d s \; \rho^{X,\;OPE}_{J^{PC}}(s) e^{-
s / M_B^2}  \;,  \label{L0}
\end{eqnarray}
and
\begin{eqnarray}
L^{X}_{J^{PC},\;1}(s_0, M_B^2) =
\frac{\partial}{\partial{\frac{1}{M_B^2}}}{L^{X}_{J^{PC},\;0}(s_0, M_B^2)} \; .
\end{eqnarray}

\section{Numerical analysis}\label{Numerical}
In the numerical calculation of QCD sum rules, the input parameters are taken from \cite{Matheus:2006xi, Cui:2011fj, Narison:2002pw,P.Col,Tang:2019nwv}:
\begin{eqnarray}
\begin{aligned}
&m_c(m_c)=\overline{m}_c=(1.275\pm0.025)\; \text{GeV}\;,&&m_b(m_b)=\overline{m}_b=(4.18\pm0.03)\; \text{GeV}   \; ,\\
& \langle \bar{q} q \rangle = - (0.24 \pm 0.01)^3 \; \text{GeV}^3 \; ,& & \langle g_s^2 G^2 \rangle = (0.88\pm0.25) \; \text{GeV}^4 \; , \\
& \langle \bar{q} g_s \sigma \cdot G q \rangle = m_0^2 \langle\bar{q} q \rangle; , & & \langle g_s^3 G^3 \rangle = (0.045 \pm 0.013) \;\text{GeV}^6 \; ,\\
& m_0^2 = (0.8 \pm 0.1) \; \text{GeV}^2\; ,
\end{aligned}
\end{eqnarray}
Here, $\overline{m}_c$ and $\overline{m}_b$ represent heavy-quark running masses in $\overline{\text{MS}}$ scheme. For light quark, the chiral quark limit mass $m_q=0$ is adopted.

Furthermore, there exist two additional parameters $s_0$ and $M_B^2$ introduced in establishing the sum rule, which can be fixed in light of the so-called standard procedures abiding by two criteria \cite{Shifman,Reinders:1984sr, P.Col}. The first one asks for the convergence of the OPE, which is to compare relative contribution of each term to the total contribution on the OPE side, and then a reliable region for $M_B^2$ will be chosen to retain the convergence. The other criterion of QCD sum rules is the pole contribution (PC). As discussed in Ref. \cite{Chen:2014vha,Azizi:2019xla,Wang:2017sto}, the large power of $s$ in the spectral density suppress the PC value, thus the pole contribution will be chosen larger than $15\%$ for hexaquark states. The two criteria can be formulated as follows:
\begin{eqnarray}
  R^{OPE}_{J^{PC}} = \left| \frac{L_{J^{PC},\;0}^{dim=10}(s_0, M_B^2)}{L_{J^{PC},\;0}(s_0, M_B^2)}\right|\, ,
\end{eqnarray}
\begin{eqnarray}
  R^{PC}_{J^{PC}} = \frac{L_{J^{PC},\;0}(s_0, M_B^2)}{L_{J^{PC},\;0}(\infty, M_B^2)} \; . \label{RatioPC}
\end{eqnarray}

To determine a proper value for $s_0$, a similar analysis in Refs. \cite{Qiao:2013dda,Tang:2016pcf,Wan:2020oxt,Finazzo:2011he,Wan:2022xkx,Zhang:2022obn} will be carried out. Therein, one needs to
find the proper value, which has an optimal window for the mass curve of the hexaquark state. Within this window, the physical quantity, that is the mass of the hexaquark state, is independent of the Borel parameter $M_B^2$ as much as possible. In practice, we will vary $\sqrt{s_0}$ by $0.2$ GeV, which obtained the lower and upper bounds hence the uncertainties of $\sqrt{s_0}$.

\begin{figure}
\includegraphics[width=6.8cm]{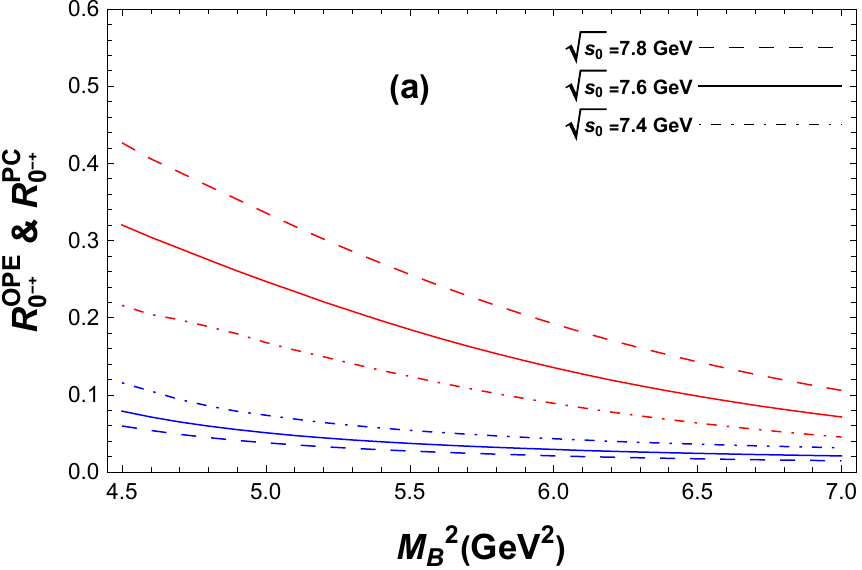}
\includegraphics[width=6.8cm]{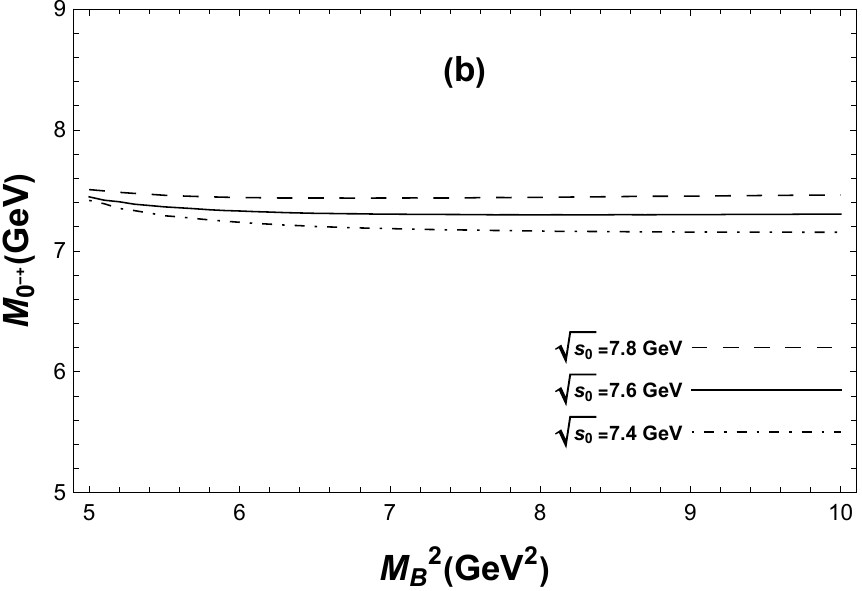}
\caption{ (a) The ratios ${R_{0^{-+}}^{OPE}}$ and ${R_{0^{-+}}^{PC}}$ as functions of the Borel parameter $M_B^2$ for different values of $\sqrt{s_0}$, where blue lines represent ${R_{0^{-+}}^{OPE}}$ and red lines denote ${R_{0^{-+}}^{PC}}$ . (b) The mass $m_{0^{-+}}$ as a function of the Borel parameter $M_B^2$ for different values of $\sqrt{s_0}$.} \label{fig0-+}
\end{figure}

With the above preparation the mass spectrum of tetracharm baryonium states will be  numerically evaluated. The rations $R^{OPE}_{0^{-+}}$ and $R^{PC}_{0^{-+}}$ are shown as functions of Borel parameter $M_B^2$ in Fig. \ref{fig0-+}(a) with different values of $\sqrt{s_0}$, i.e., $7.4$, $7.6$ and $7.8$ GeV. The dependence relationships between $m_{0^{-+}}$ and parameter $M_B^2$ are given in Fig. \ref{fig0-+}(b). The optimal Borel window is found in range $5.3 \le M_B^2 \le 6.3\; \text{GeV}^2$, and the mass $m_{0^{-+}}$ can be extracted as follow:
\begin{eqnarray}
m_{0^{-+}} &=& (7.33\pm 0.12)\; \text{GeV}\;.\label{m1}
\end{eqnarray}

\begin{figure}
\includegraphics[width=6.8cm]{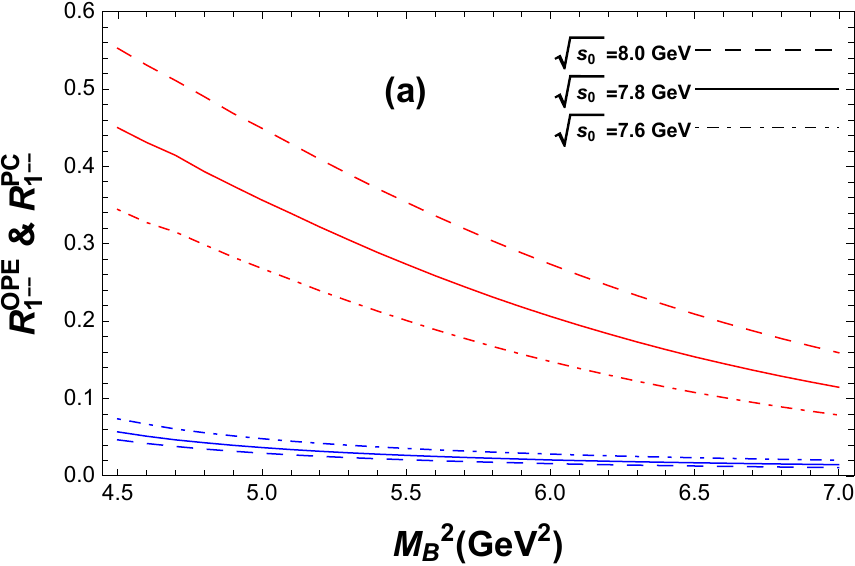}
\includegraphics[width=6.8cm]{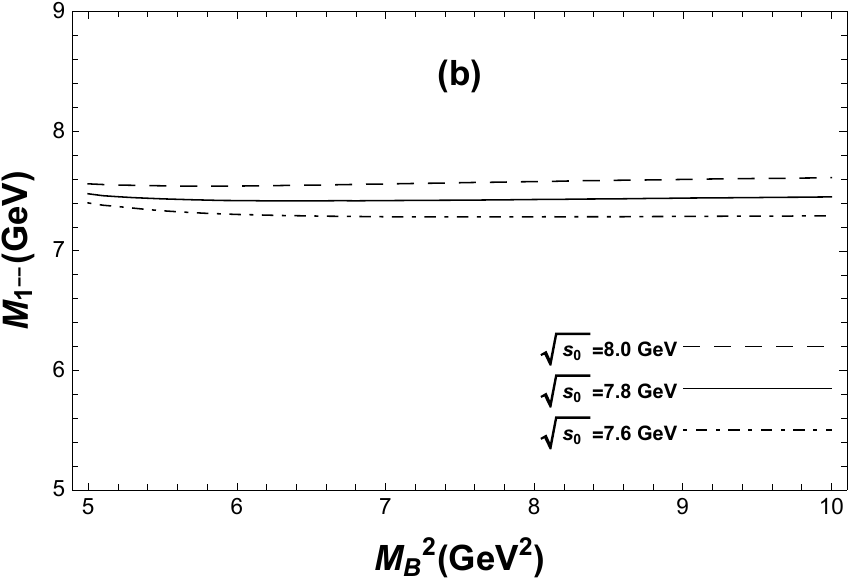}
\caption{The same caption as in Fig \ref{fig0-+}, but for the quantum number of $1^{--}$.} \label{fig1--}
\end{figure}

\begin{figure}
\includegraphics[width=6.8cm]{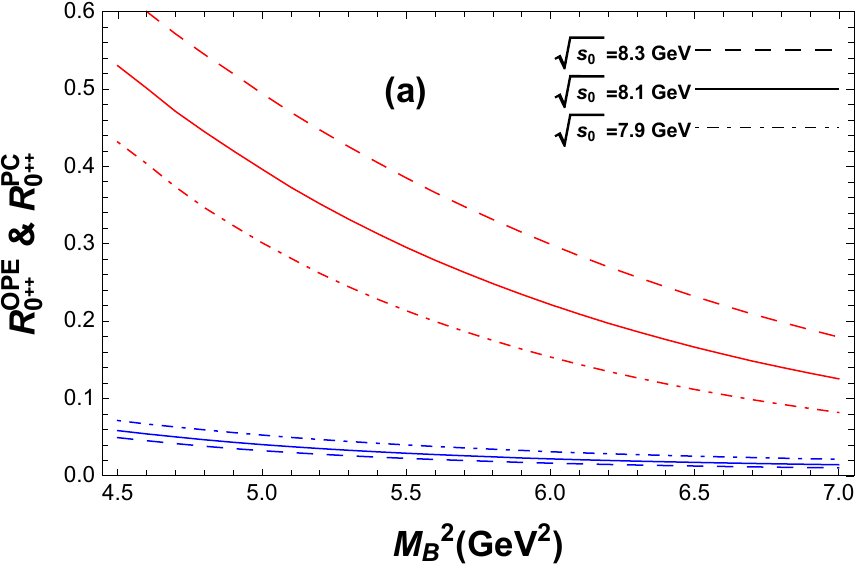}
\includegraphics[width=6.8cm]{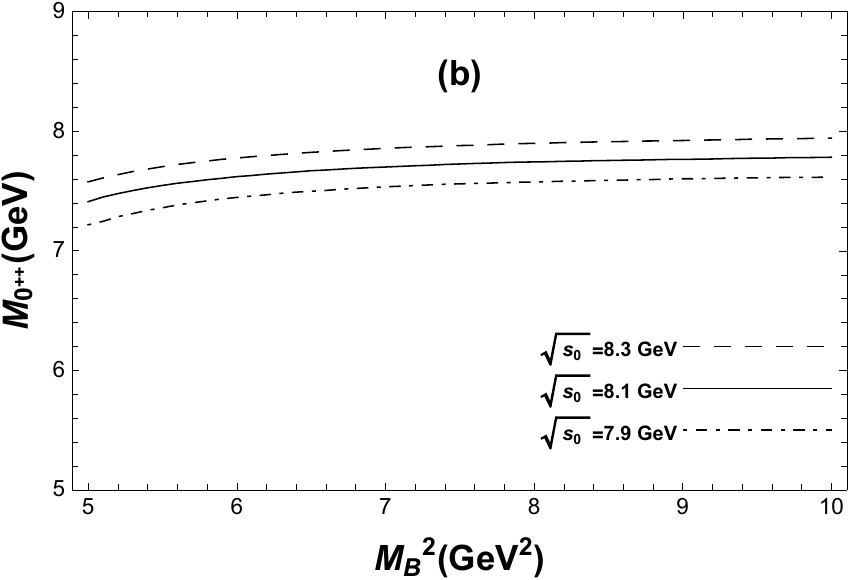}
\caption{The same caption as in Fig \ref{fig0-+}, but for the  quantum number of $0^{++}$.} \label{fig0++}
\end{figure}

\begin{figure}
\includegraphics[width=6.8cm]{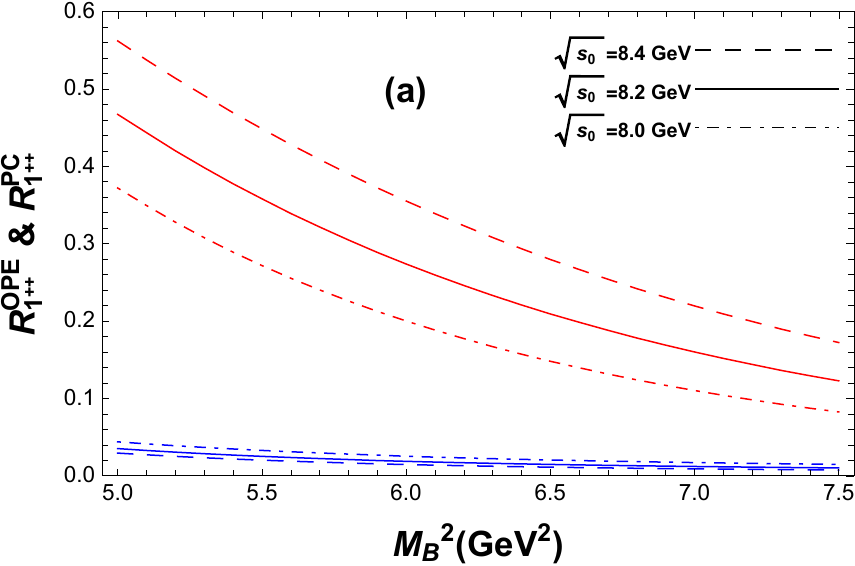}
\includegraphics[width=6.8cm]{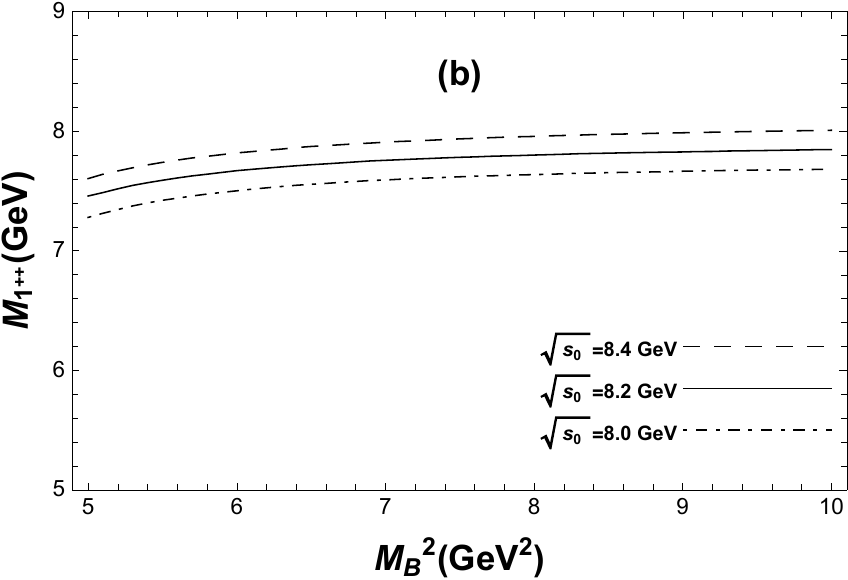}
\caption{The same caption as in Fig \ref{fig0-+}, but for the quantum number of $1^{++}$.} \label{fig1++}
\end{figure}

With the same analyses, and the OPE, pole contribution and masses as functions of Borel parameter $M_B^2$ can be found in Figs.~\ref{fig1--}$-$\ref{fig1++}, respectively, the masses $m_{1^{--}}$, $m_{0^{++}}$ and $m_{1^{++}}$ can be extracted as follow:
\begin{eqnarray}
m_{1^{--}} &=& (7.42\pm 0.13)\; \text{GeV}\;,\\\label{m2}
m_{0^{++}} &=& (7.68\pm 0.17)\; \text{GeV}\;,\\\label{m3}
m_{1^{++}} &=& (7.76\pm 0.12)\; \text{GeV}\;.\label{m4}
\end{eqnarray}

\begin{figure}
\includegraphics[width=6.8cm]{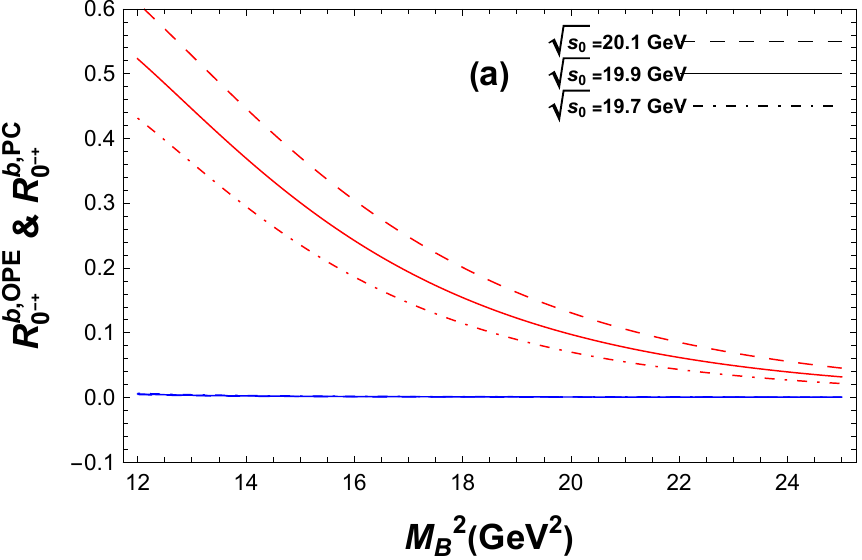}
\includegraphics[width=6.8cm]{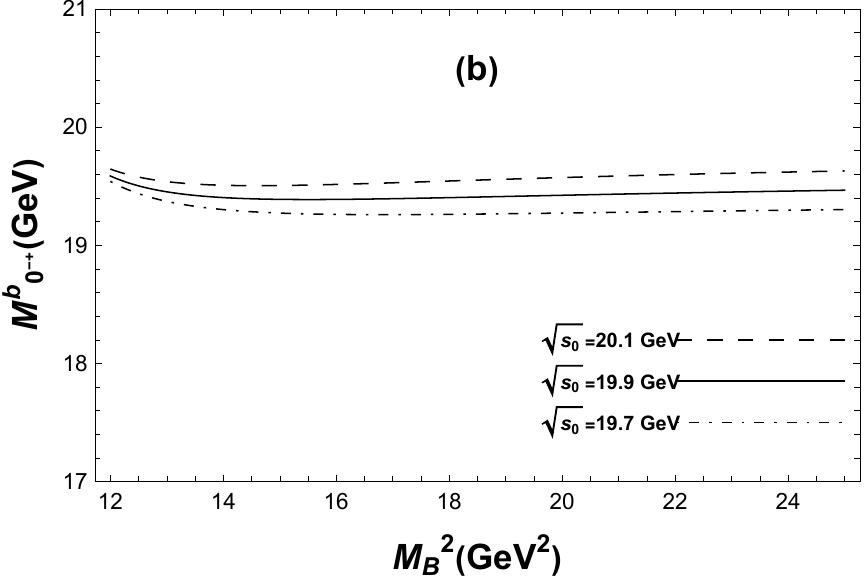}
\caption{ The same caption as in Fig \ref{fig0-+}, but for the tetra-bottom baryonium for the quantum number of $0^{-+}$.} \label{figb0-+}
\end{figure}

\begin{figure}
\includegraphics[width=6.8cm]{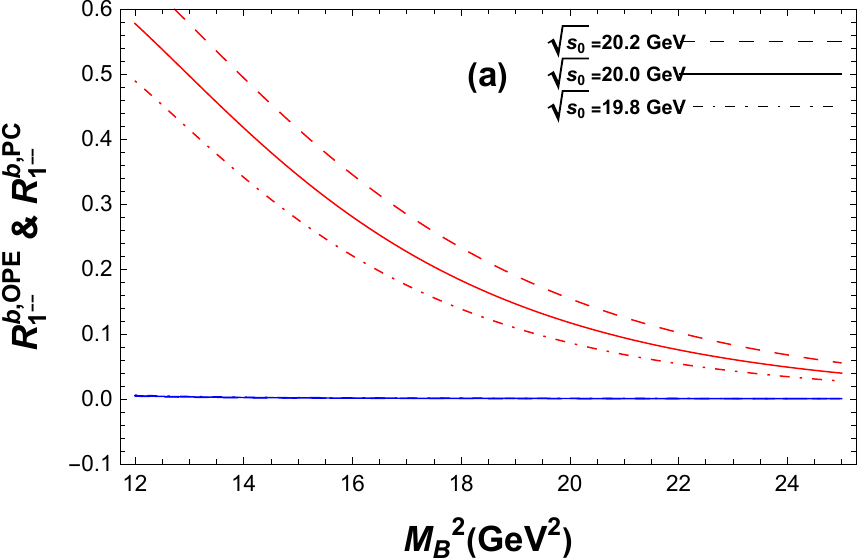}
\includegraphics[width=6.8cm]{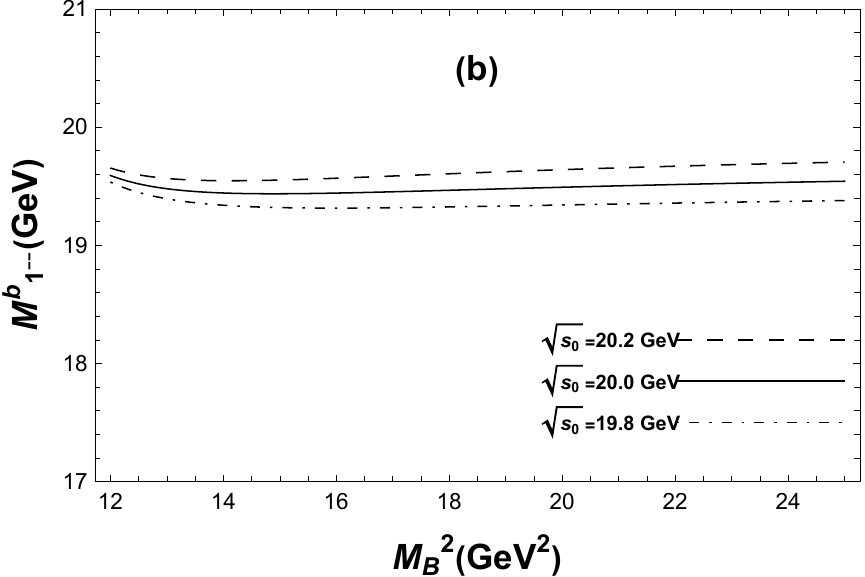}
\caption{ The same caption as in Fig \ref{fig0-+}, but for the tetra-bottom baryonium for the quantum number of $1^{--}$.} \label{figb1--}
\end{figure}

\begin{figure}
\includegraphics[width=6.8cm]{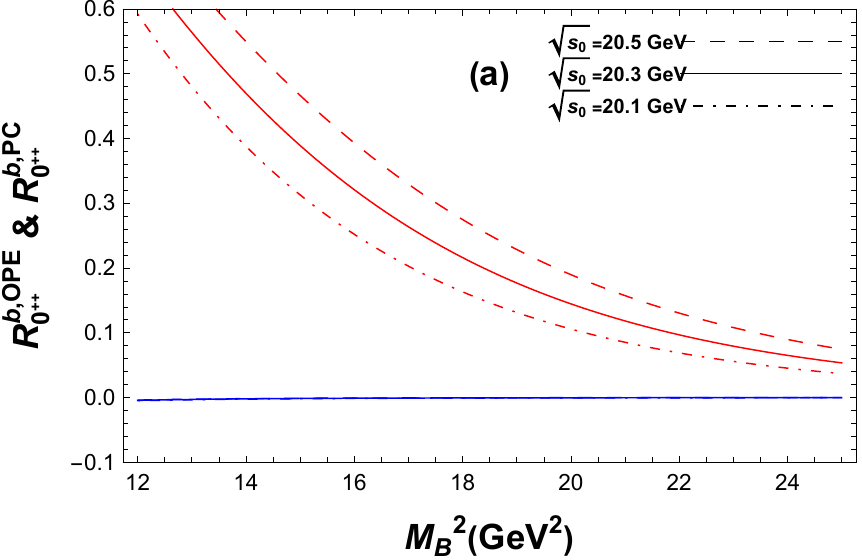}
\includegraphics[width=6.8cm]{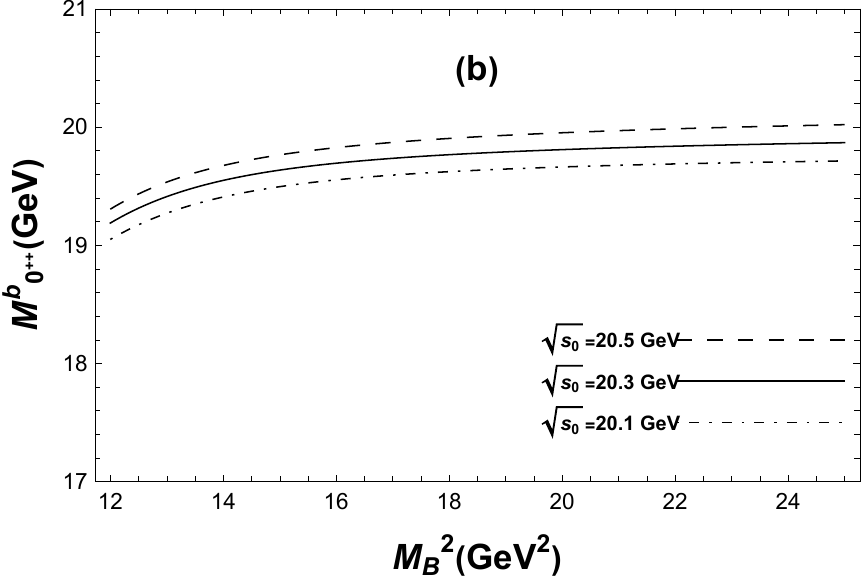}
\caption{ The same caption as in Fig \ref{fig0-+}, but for the tetra-bottom baryonium for the quantum number of $0^{++}$.} \label{figb0++}
\end{figure}

\begin{figure}
\includegraphics[width=6.8cm]{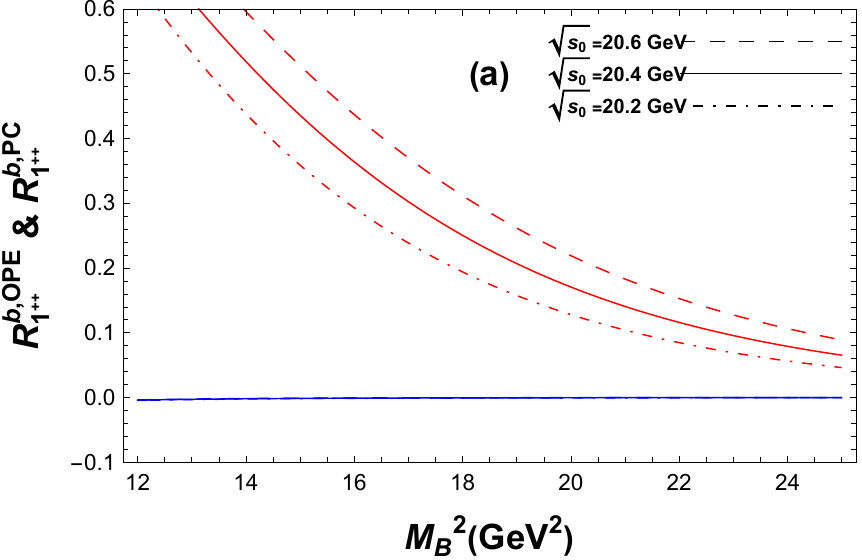}
\includegraphics[width=6.8cm]{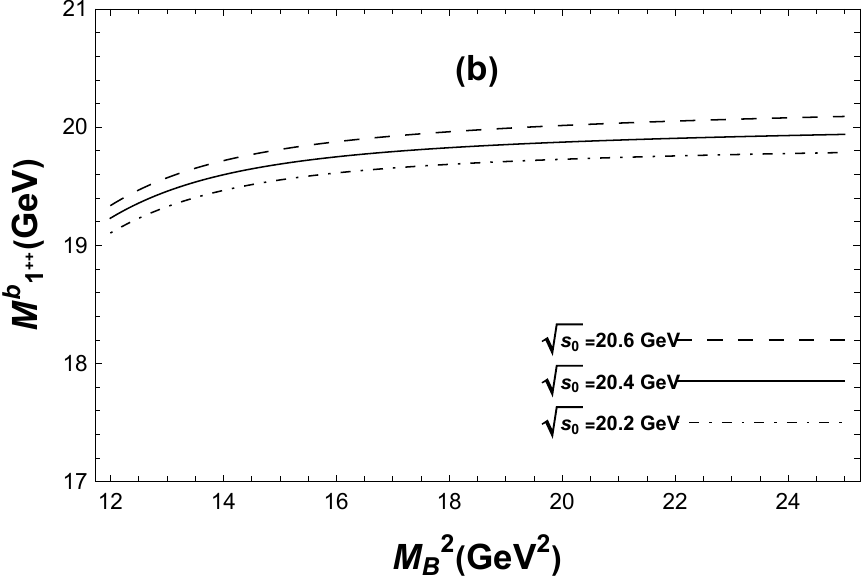}
\caption{ The same caption as in Fig \ref{fig0-+}, but for the tetra-bottom baryonium for the quantum number of $1^{++}$.} \label{figb1++}
\end{figure}

\begin{figure}
\includegraphics[width=6.8cm]{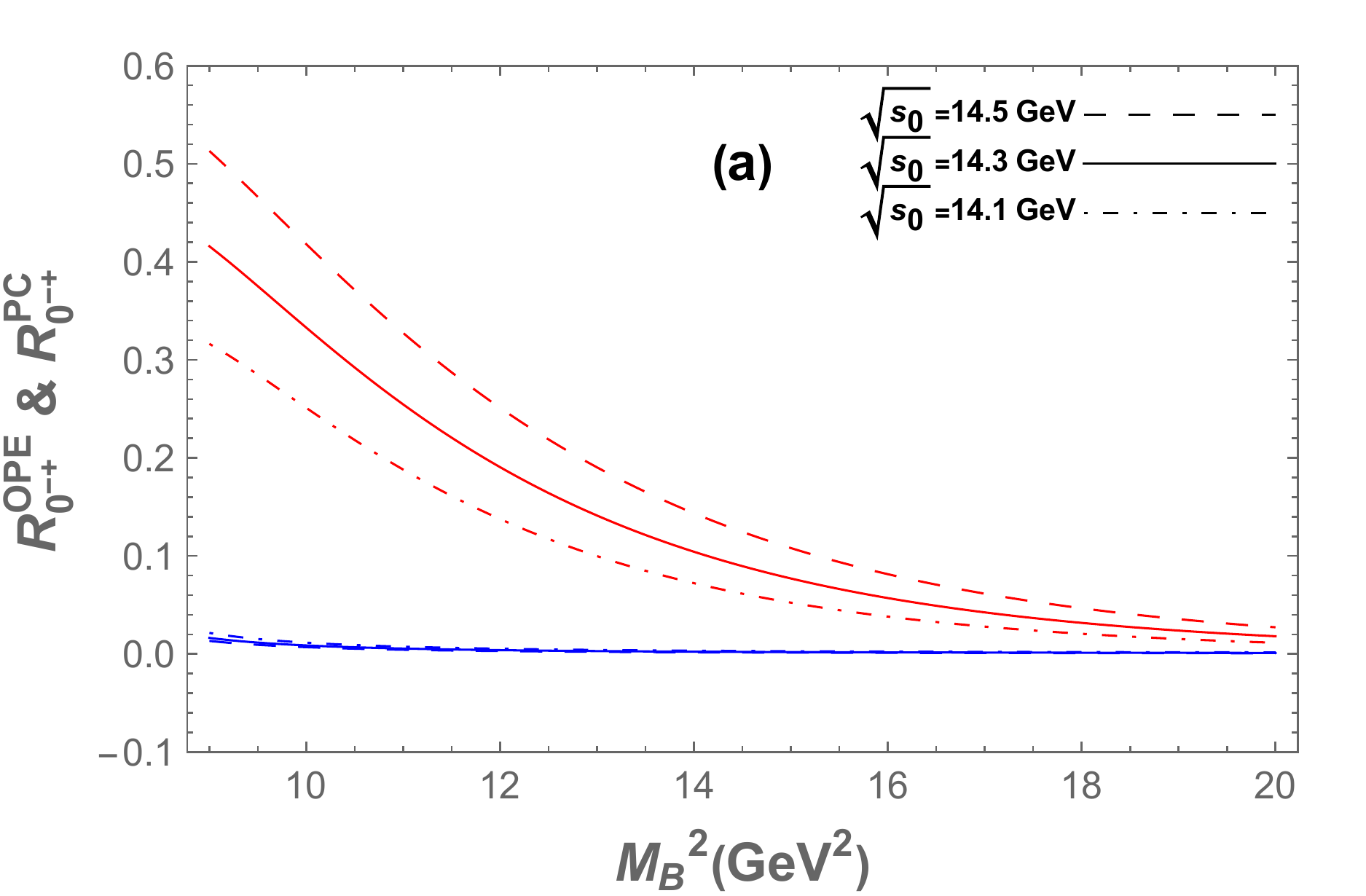}
\includegraphics[width=6.8cm]{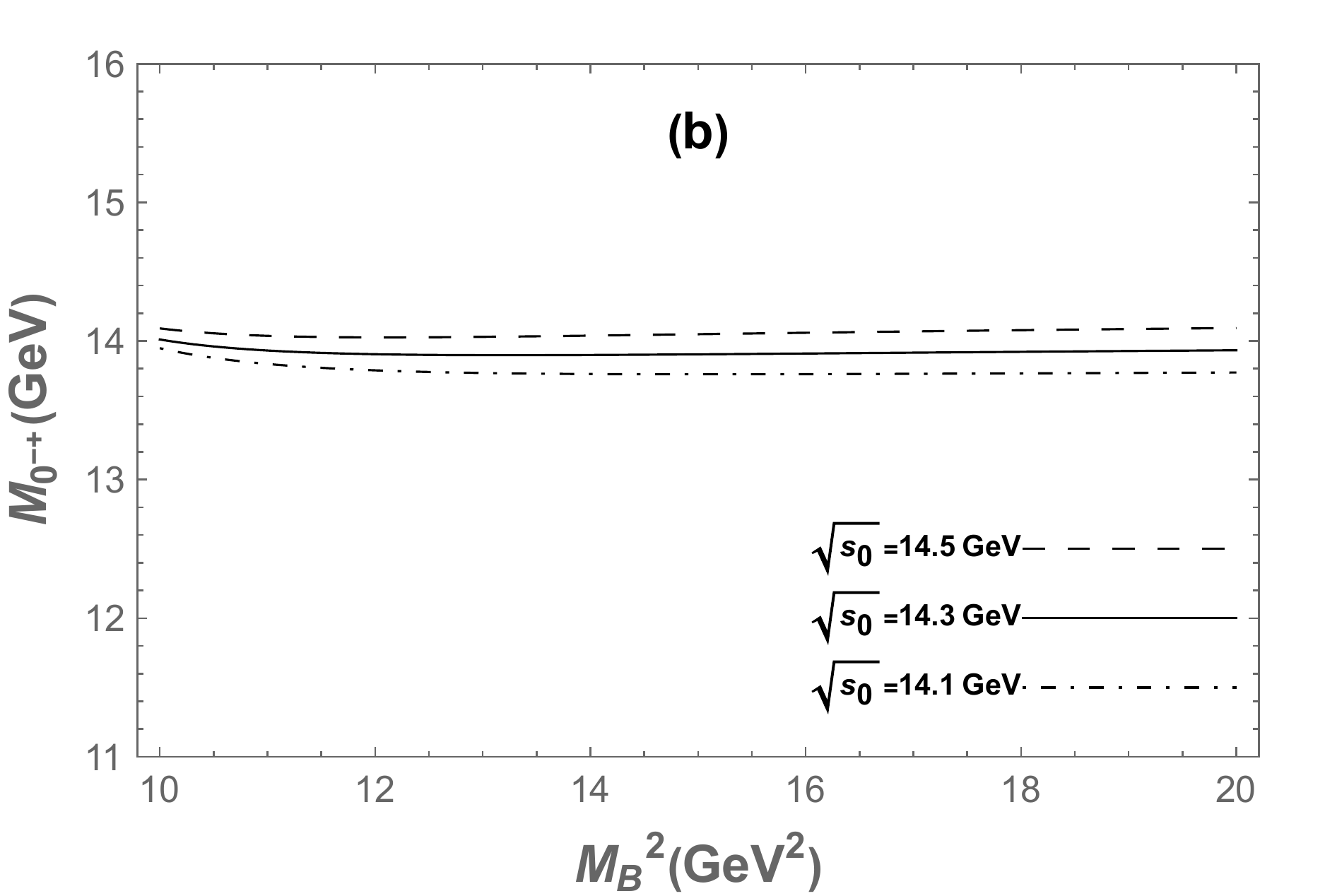}
\caption{The same caption as in Fig \ref{fig0-+}, but for the $\bar{\Xi}_{bc}\Xi_{bc}$ baryonium state with the quantum number of $0^{-+}$.} \label{figbcbc0-+}
\end{figure}

\begin{figure}
\includegraphics[width=6.8cm]{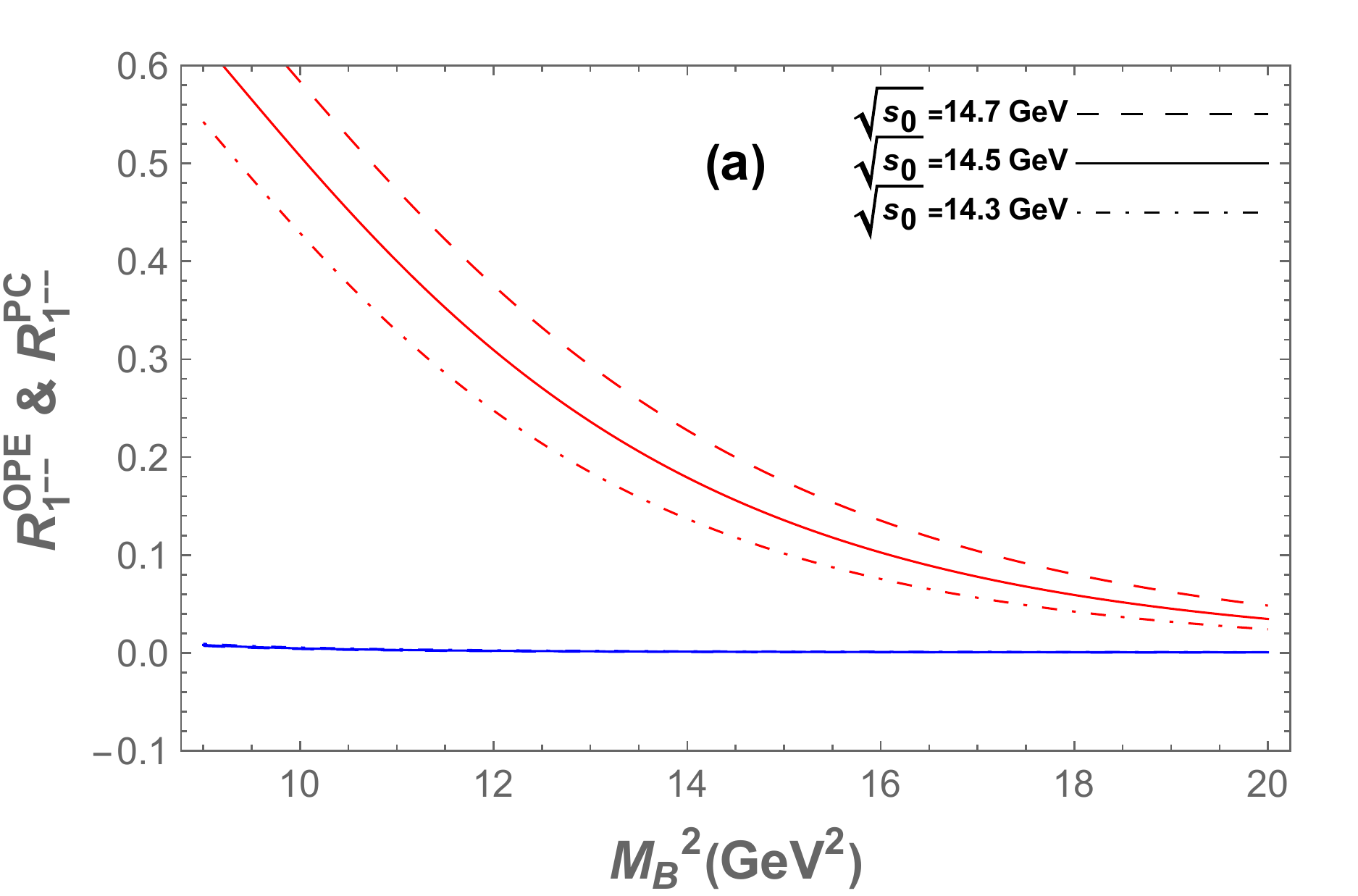}
\includegraphics[width=6.8cm]{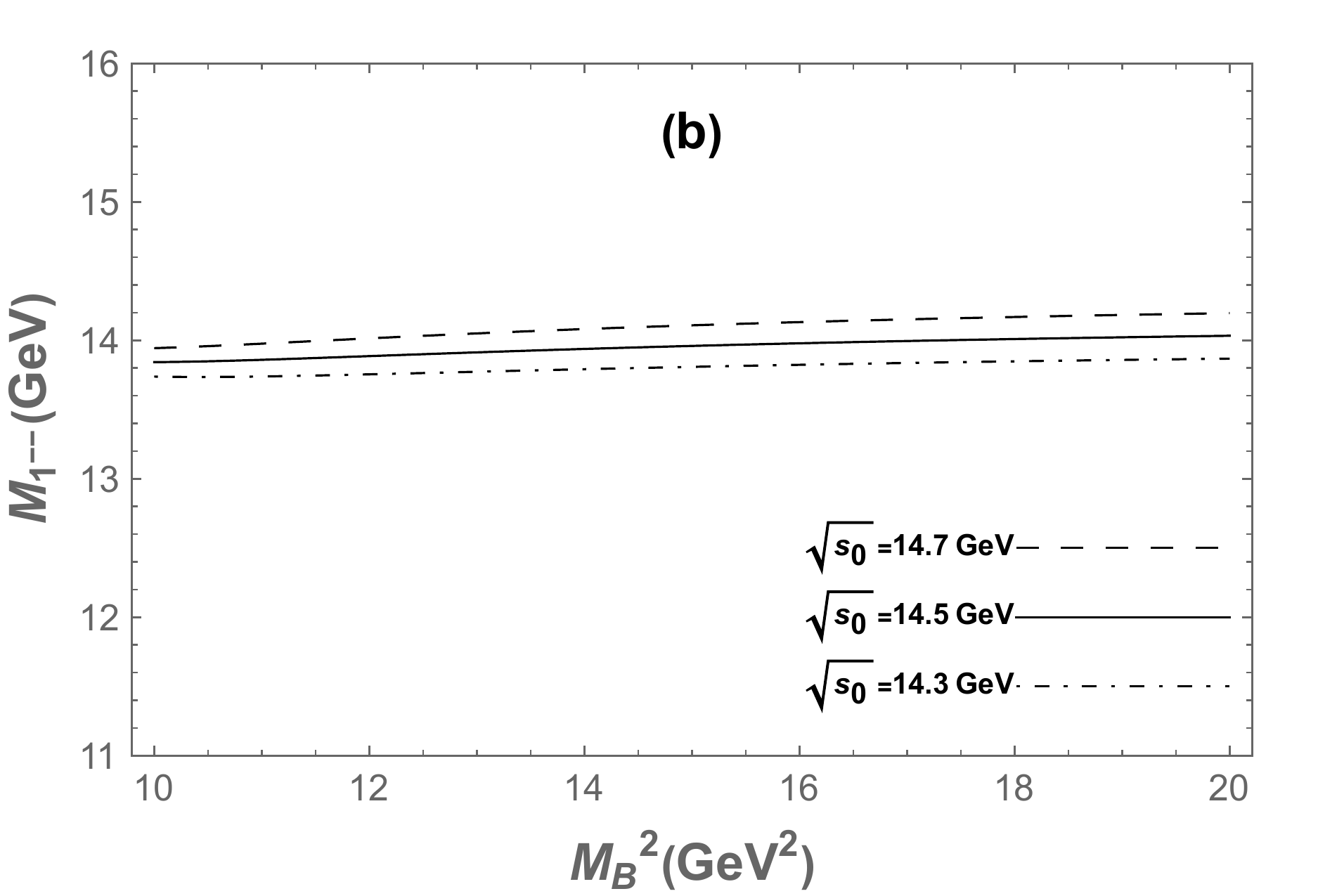}
\caption{The same caption as in Fig \ref{fig0-+}, but for the $\bar{\Xi}_{bc}\Xi_{bc}$ baryonium state with the quantum number of $1^{--}$.} \label{figbcbc1--}
\end{figure}

\begin{figure}
\includegraphics[width=6.8cm]{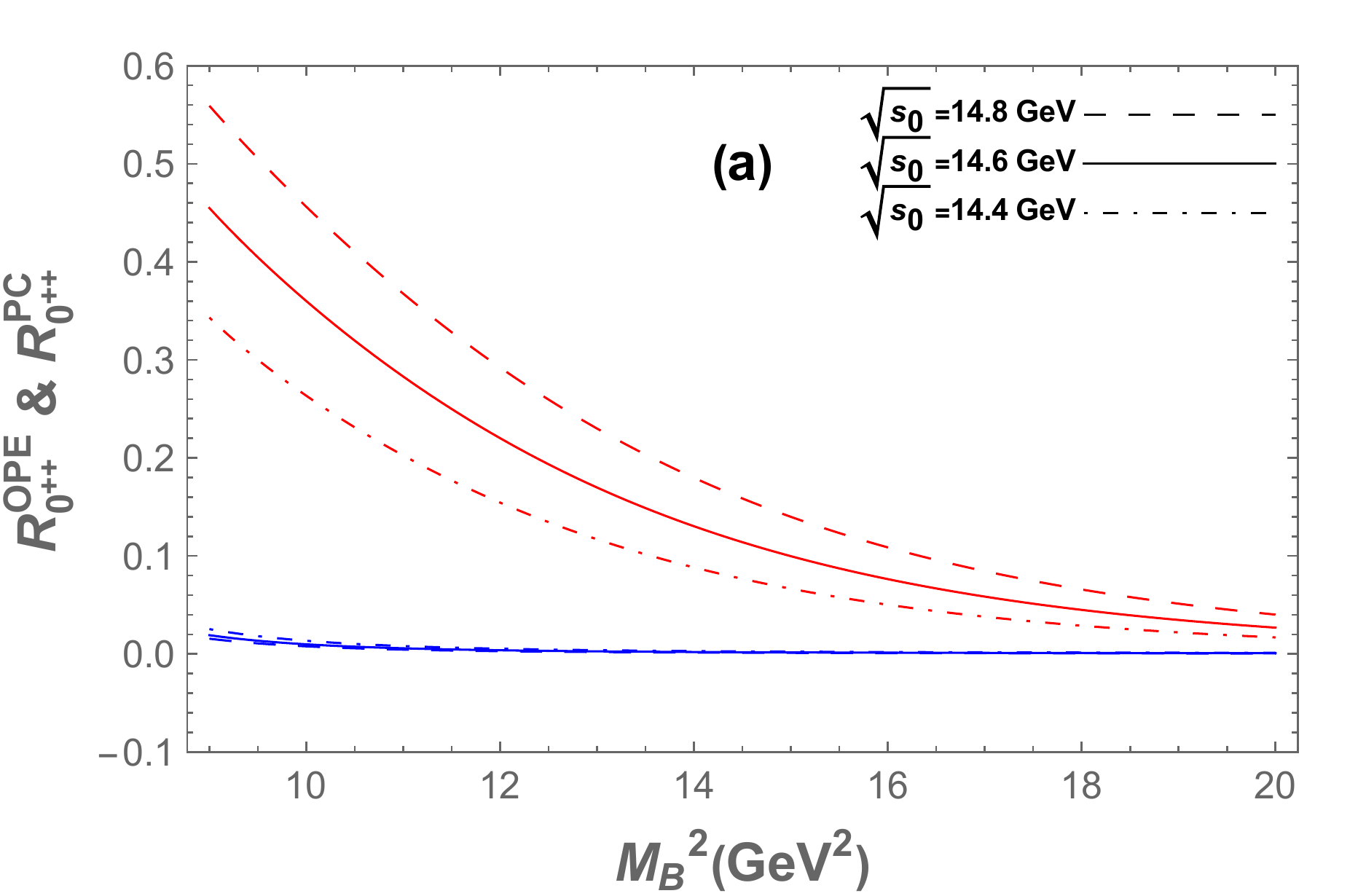}
\includegraphics[width=6.8cm]{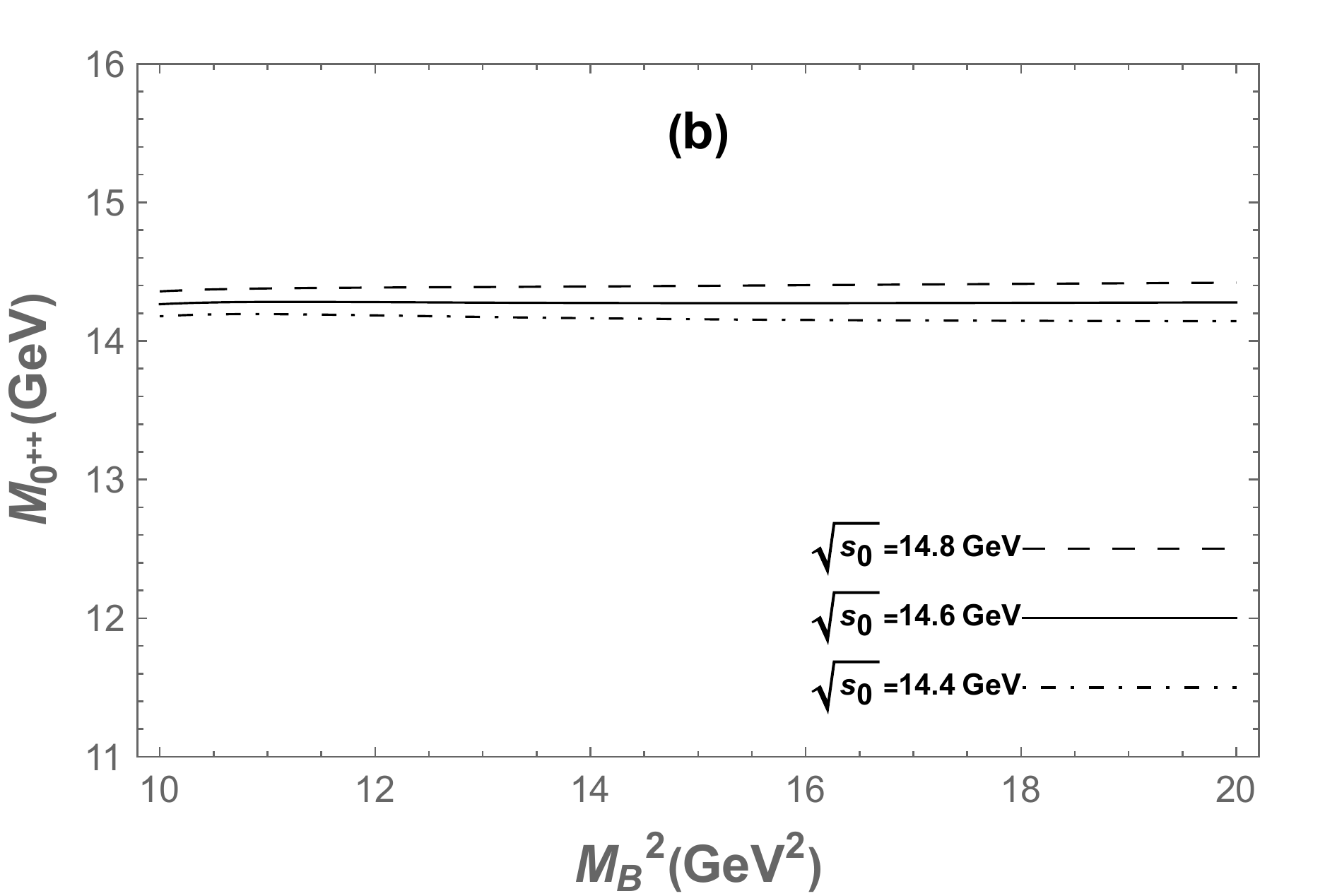}
\caption{The same caption as in Fig \ref{fig0-+}, but for the $\bar{\Xi}_{bc}\Xi_{bc}$ baryonium state with the quantum number of $0^{++}$.} \label{figbcbc0++}
\end{figure}

\begin{figure}
\includegraphics[width=6.8cm]{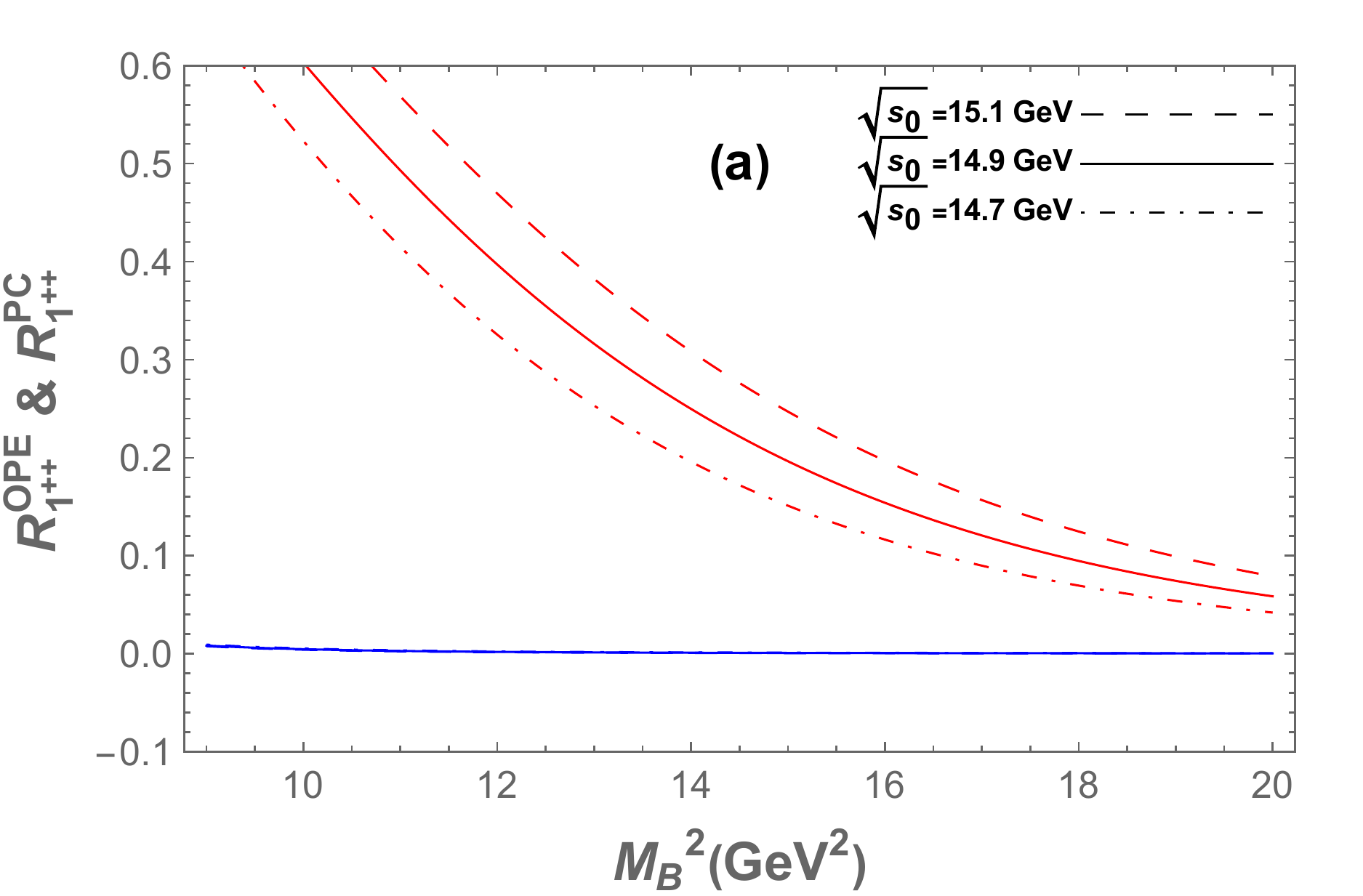}
\includegraphics[width=6.8cm]{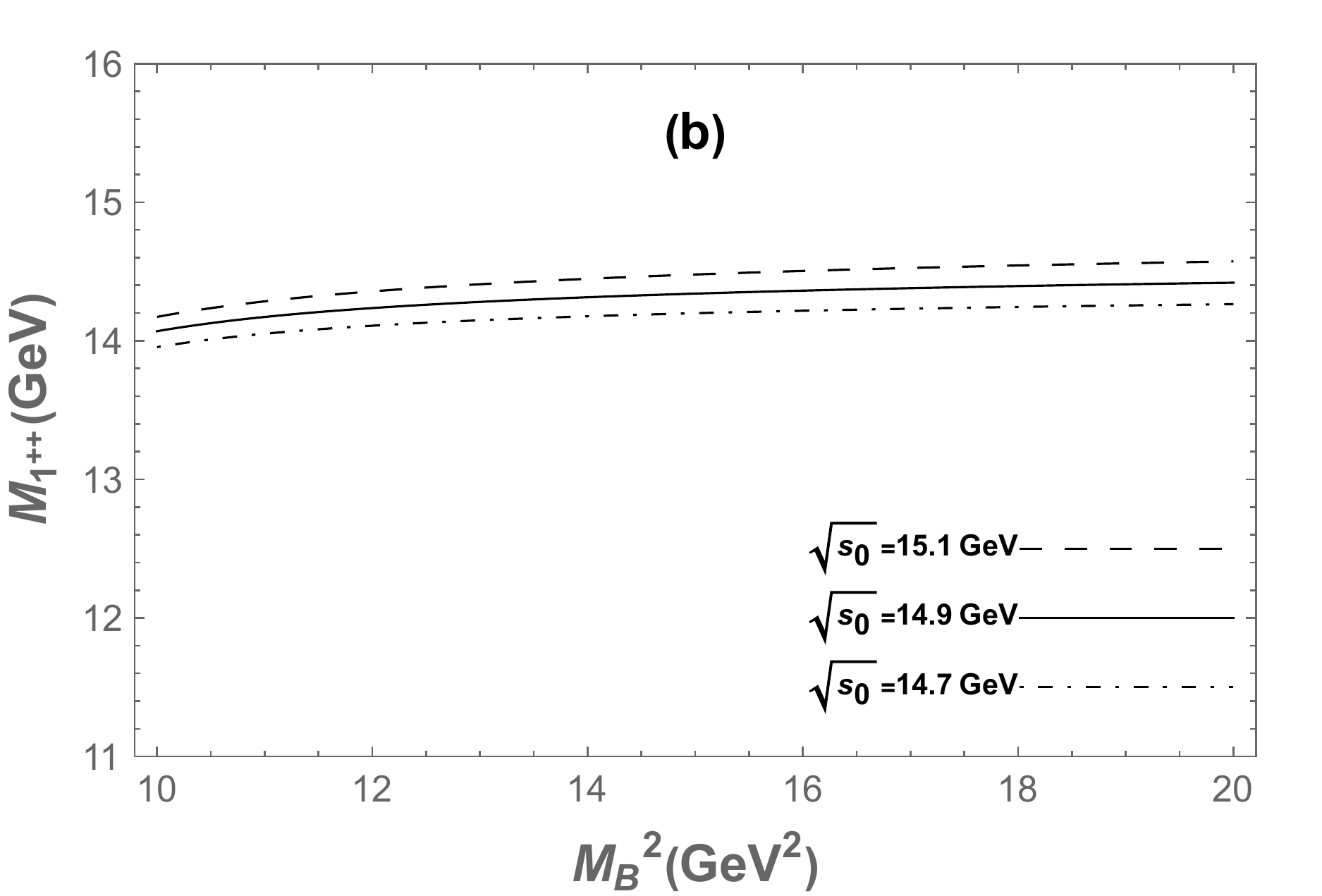}
\caption{The same caption as in Fig \ref{fig0-+}, but for the $\bar{\Xi}_{bc}\Xi_{bc}$ baryonium state with the quantum number of $1^{++}$.} \label{figbcbc1++}
\end{figure}

\begin{figure}
\includegraphics[width=6.8cm]{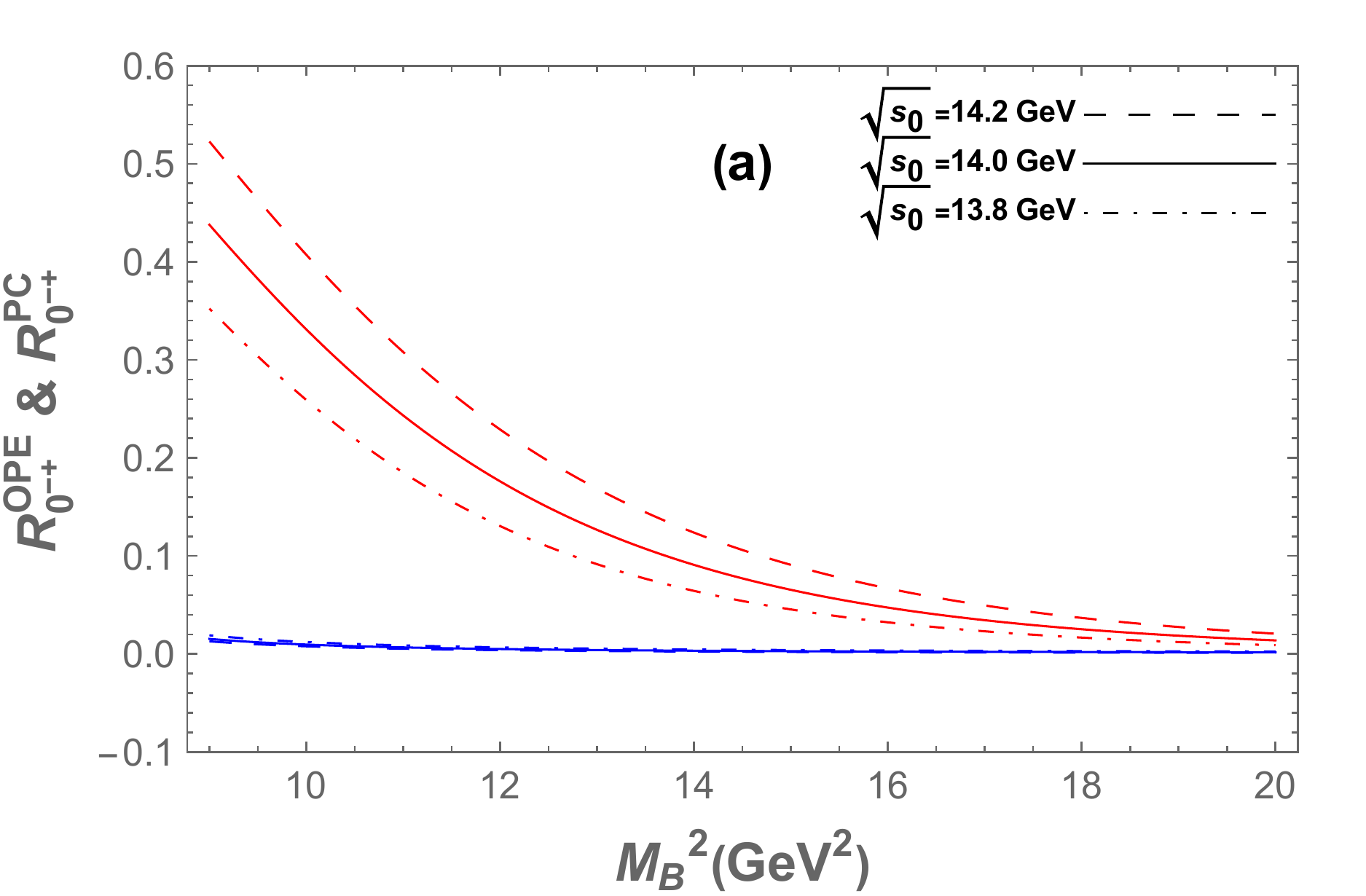}
\includegraphics[width=6.8cm]{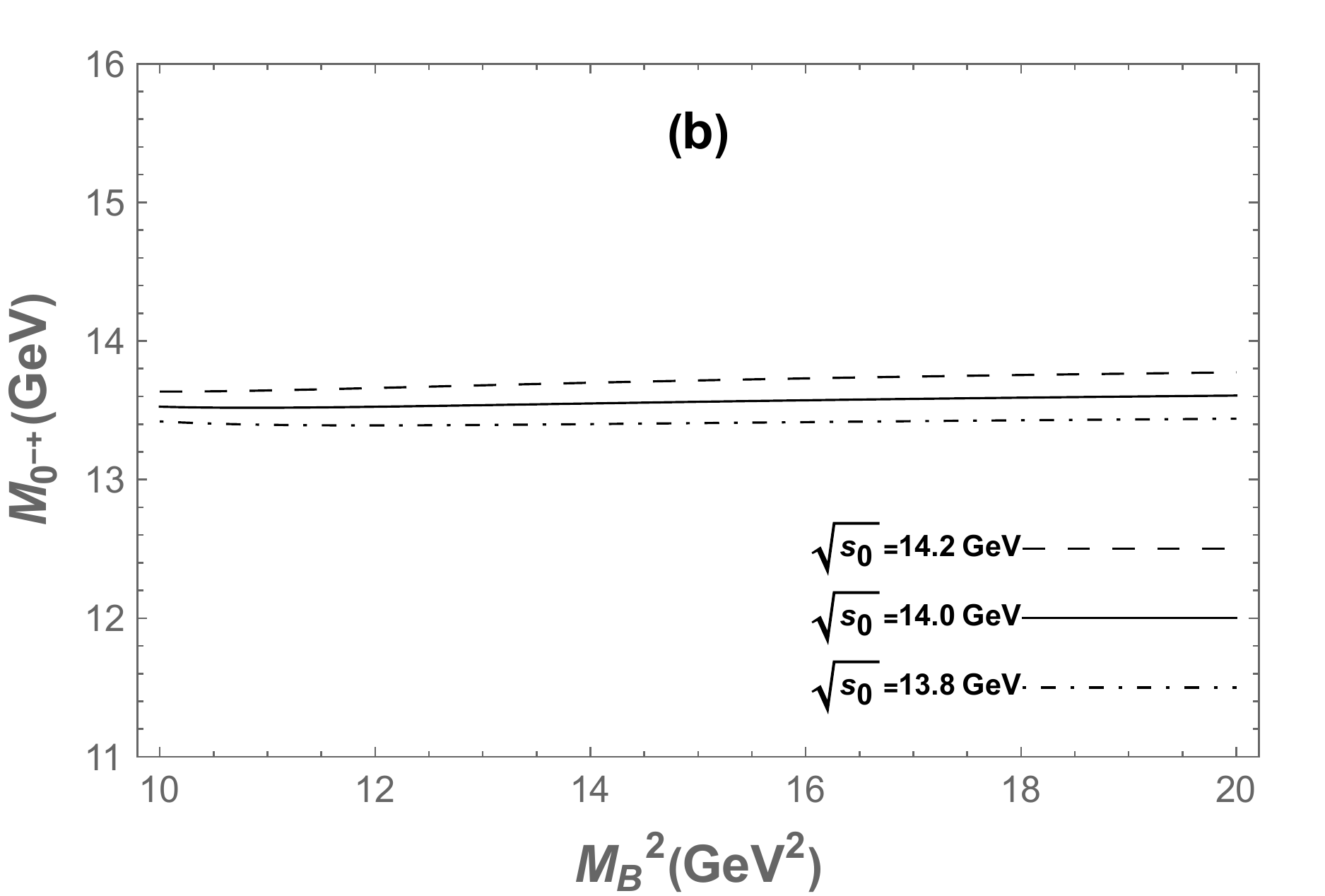}
\caption{The same caption as in Fig \ref{fig0-+}, but for the $\bar{\Xi}_{cc}\Xi_{bb}$ baryonium state with the quantum number of $0^{-+}$.} \label{figbbcc0-+}
\end{figure}

\begin{figure}
\includegraphics[width=6.8cm]{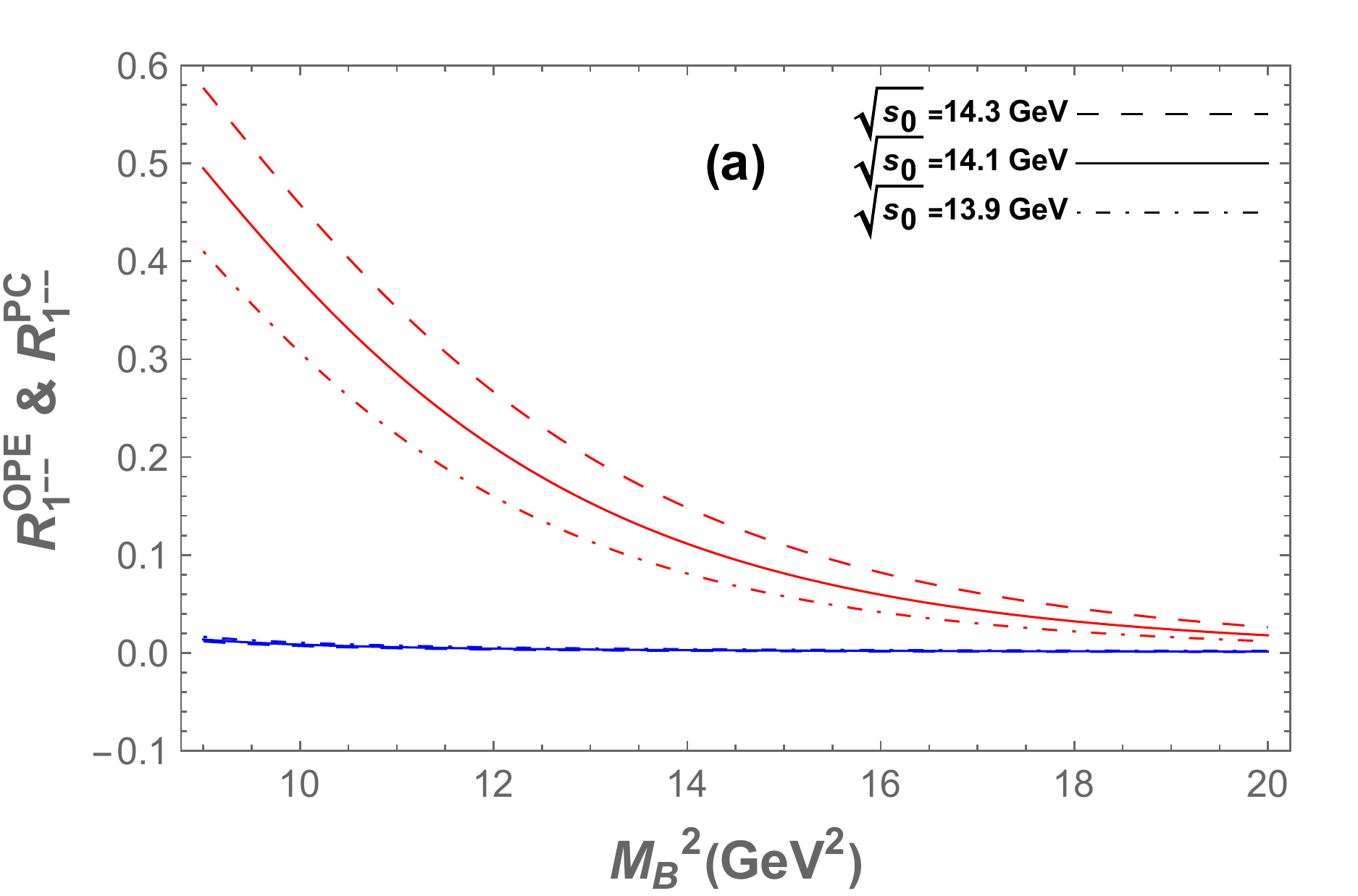}
\includegraphics[width=6.8cm]{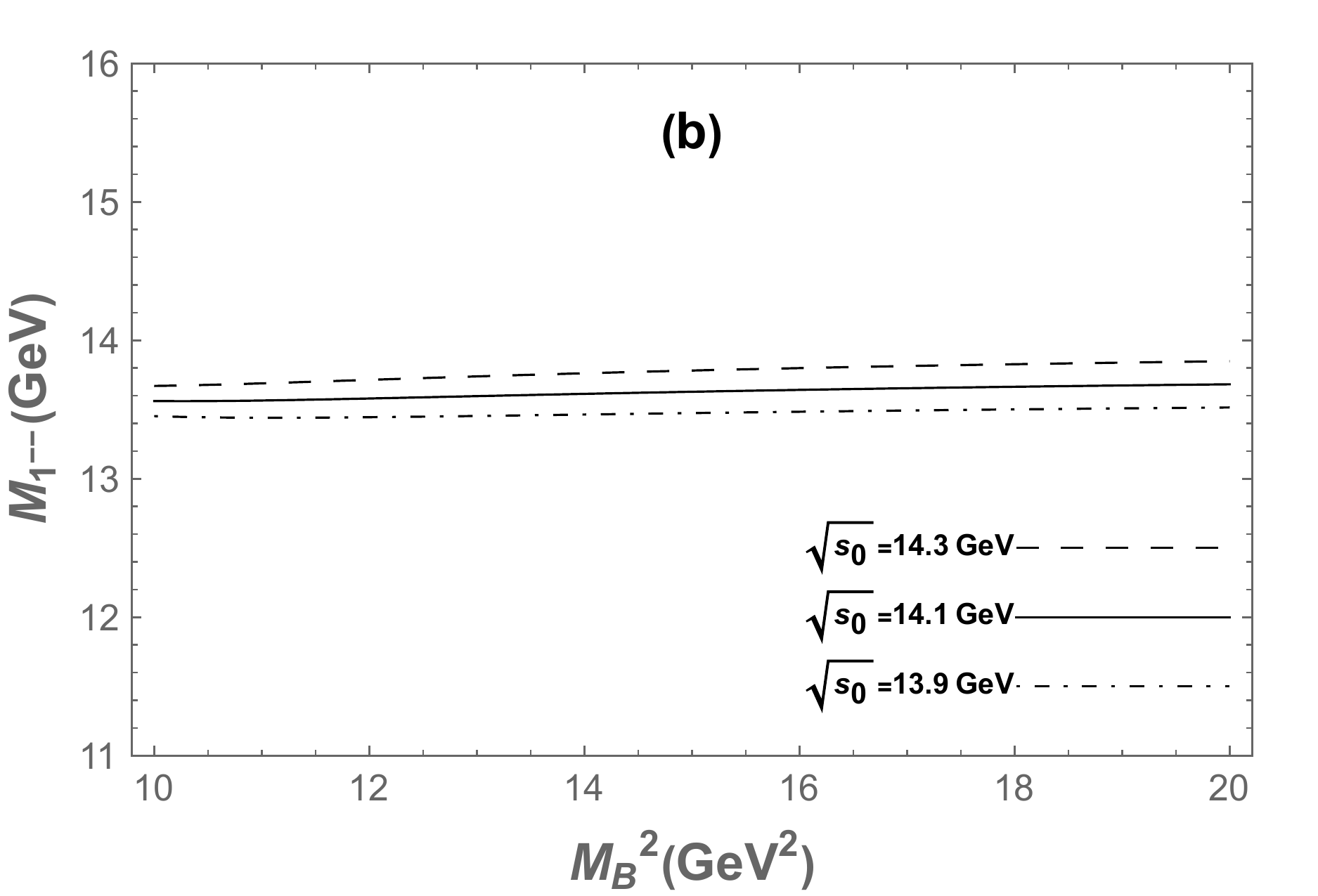}
\caption{The same caption as in Fig \ref{fig0-+}, but for the $\bar{\Xi}_{cc}\Xi_{bb}$ baryonium state with the quantum number of $1^{--}$.} \label{figbbcc1--}
\end{figure}

\begin{figure}
\includegraphics[width=6.8cm]{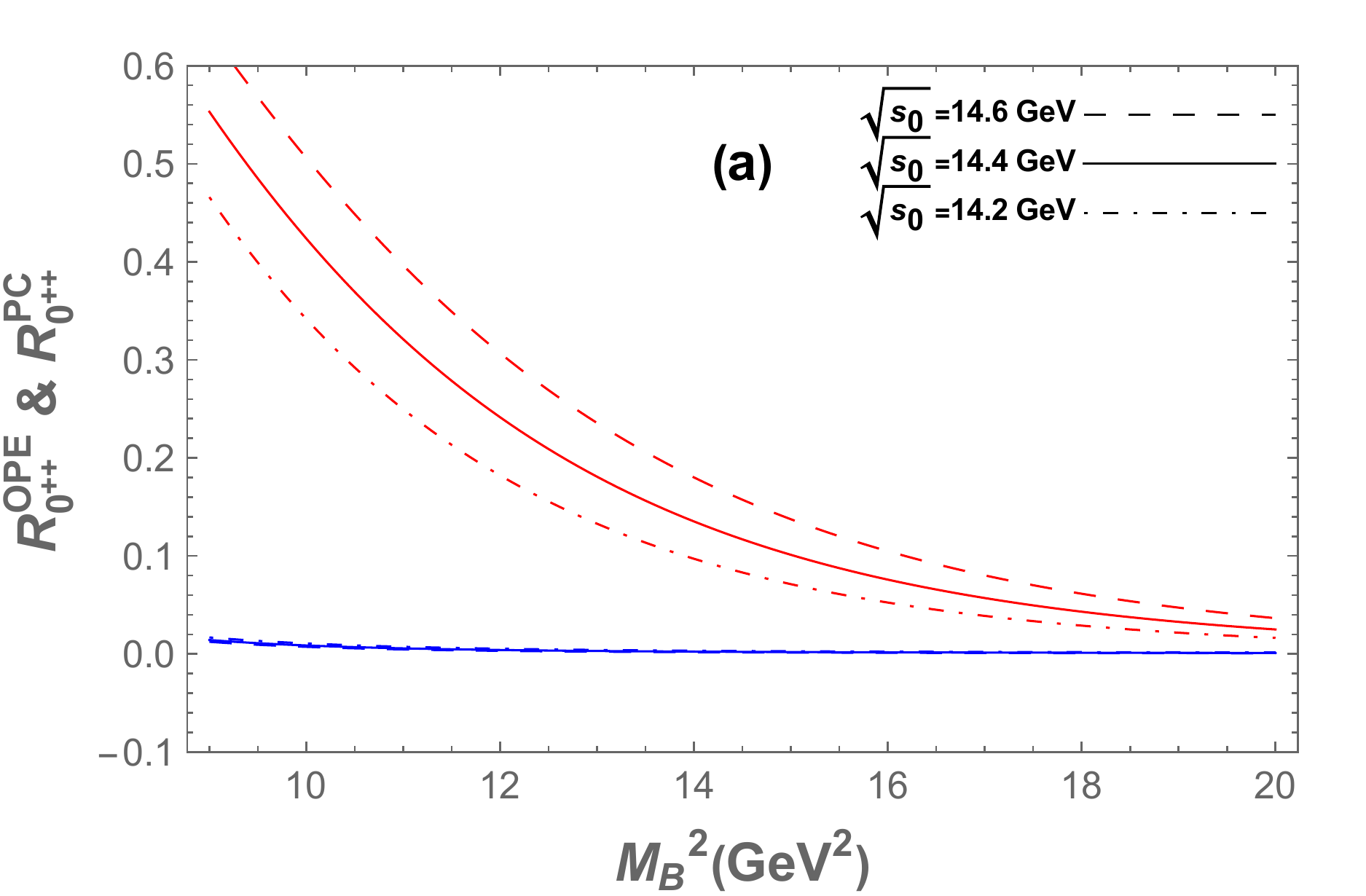}
\includegraphics[width=6.8cm]{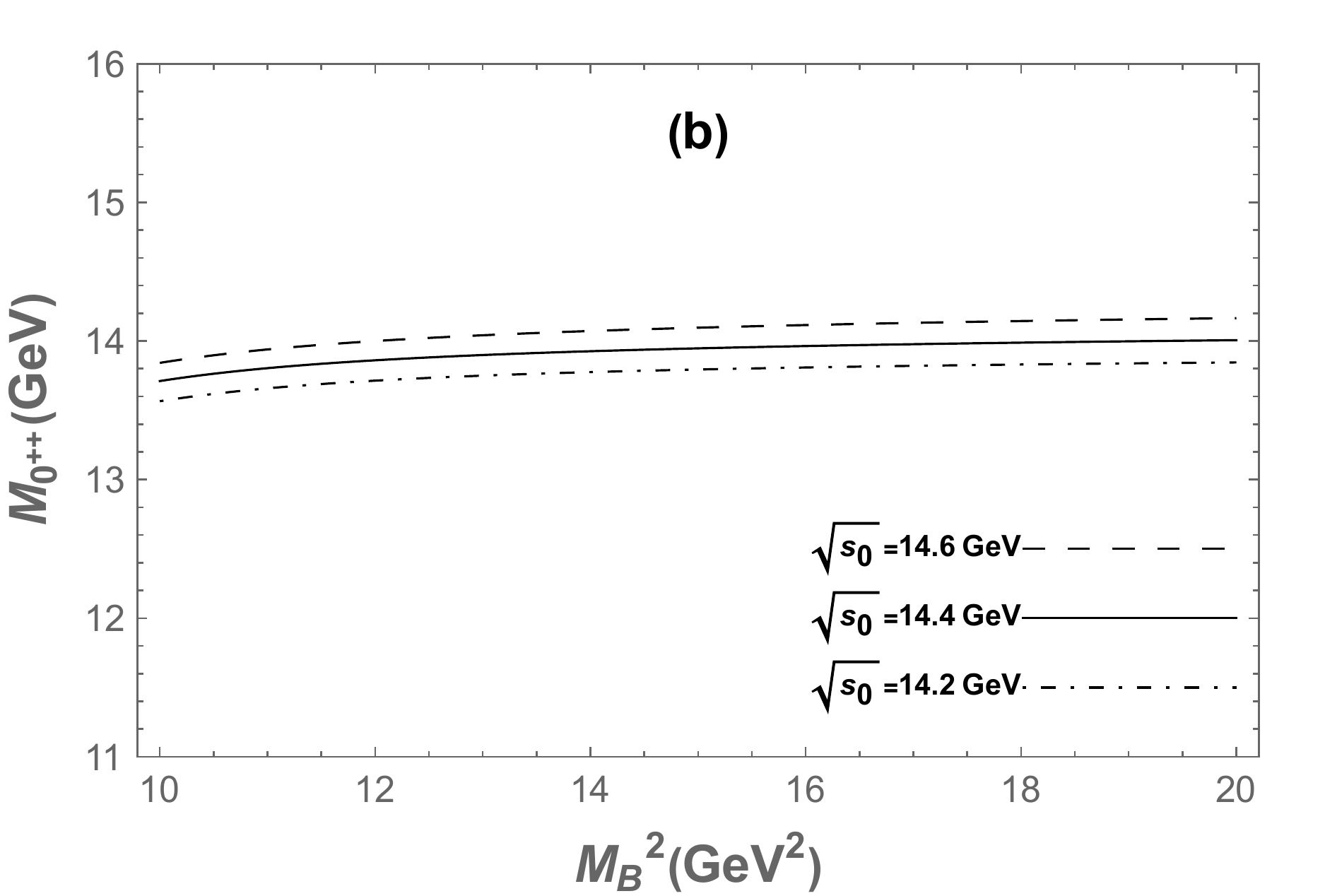}
\caption{The same caption as in Fig \ref{fig0-+}, but for the $\bar{\Xi}_{cc}\Xi_{bb}$ baryonium state with the quantum number of $0^{++}$.} \label{figbbcc0++}
\end{figure}

\begin{figure}
\includegraphics[width=6.8cm]{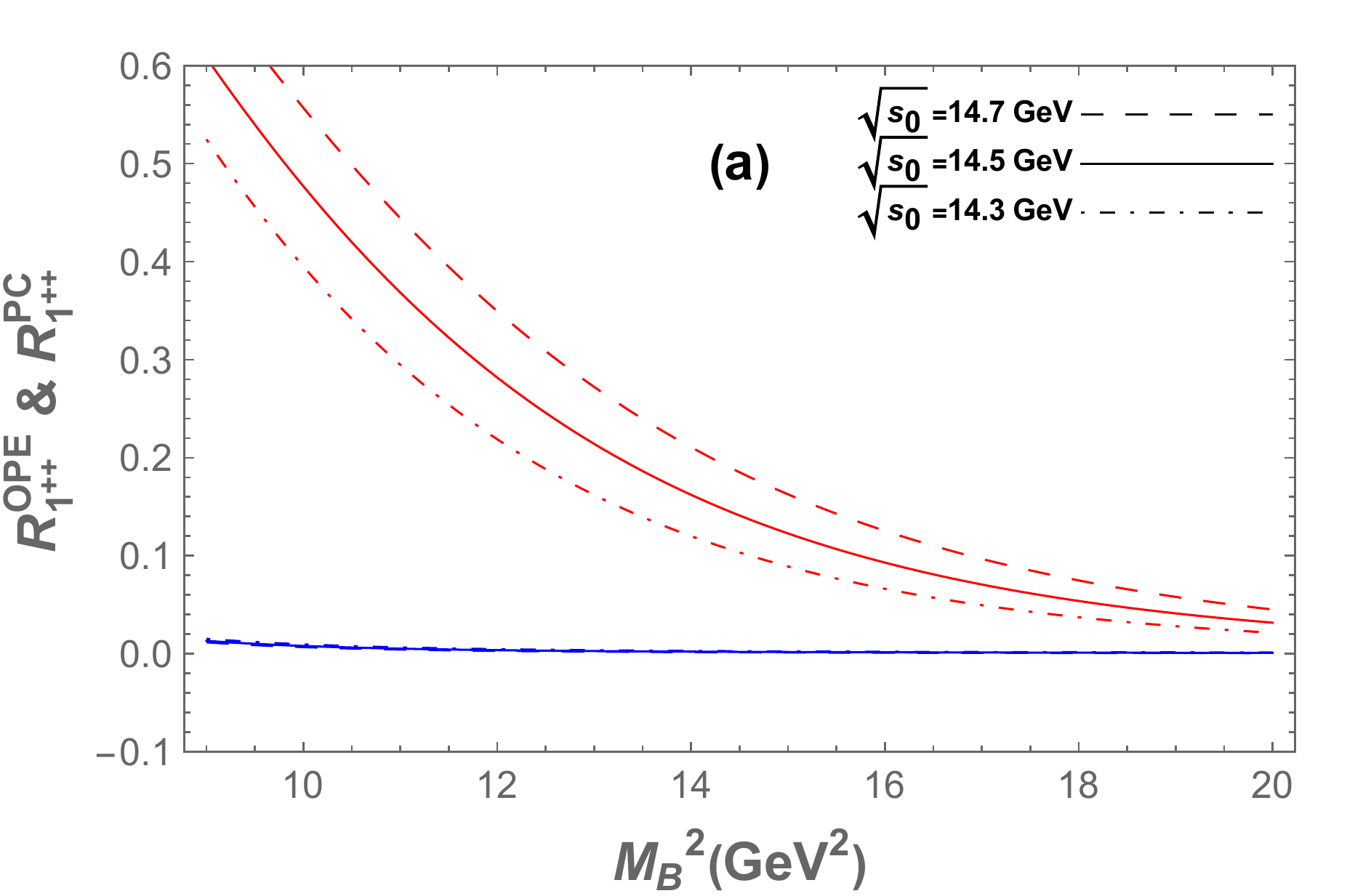}
\includegraphics[width=6.8cm]{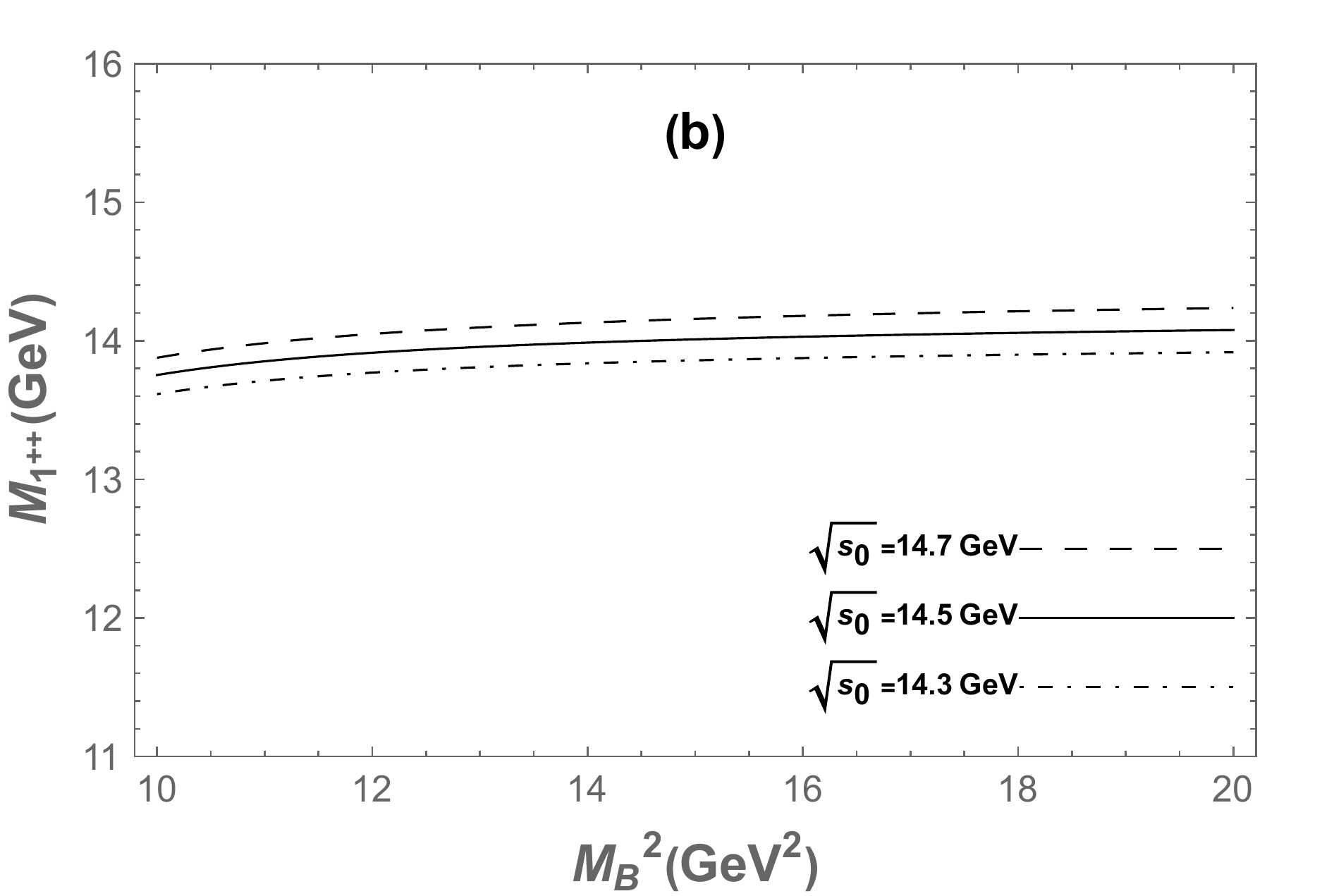}
\caption{The same caption as in Fig \ref{fig0-+}, but for the $\bar{\Xi}_{cc}\Xi_{bb}$ baryonium state with the quantum number of $1^{++}$.} \label{figbbcc1++}
\end{figure}

Similarly, we can evaluate the tetrabottom baryonium states. By using the obtained analytical results but with $m_c$ being replaced by $m_b$, the corresponding masses are readily obtained, that is
\begin{eqnarray}
m^b_{0^{-+}} &=& (19.41\pm 0.15)\; \text{GeV}\;,\\\label{m5}
m^b_{1^{--}} &=& (19.48\pm 0.15)\; \text{GeV}\;,\\\label{m6}
m^b_{0^{++}} &=& (19.77\pm 0.19)\; \text{GeV}\;,\\\label{m7}
m^b_{1^{++}} &=& (19.84\pm 0.19)\; \text{GeV}\;.\label{m8}
\end{eqnarray}
Here, superscript $b$ denotes the $b$-sector baryonium states. The convergence of the OPE, pole contribution and the masses of tetrabottom baryonium states are shown as functions of Borel parameter $M_B^2$ in Figs.~\ref{figb0-+}$-$\ref{figb1++} for the quantum number of $0^{-+}$, $1^{--}$, $0^{++}$ and $1^{++}$, respectively.

The errors of the results, (\ref{m1})$-$(\ref{m8}), mainly stem from the uncertainties in quark masses, condensates and threshold parameter $\sqrt{s_0}$. For the convenience of reference, a collection of continuum thresholds, Borel parameters, and predicted masses of tetracharm and tetrabottom baryonium states are tabulated in Table \ref{mass_baryonium}. Additionally, the spectra of the $\bar{\Xi}_{bc}\Xi_{bc}$ and $\bar{\Xi}_{cc}\Xi_{bb}$ baryonium states are also analyzed and tabulated in Table \ref{mass_baryonium}. The convergence of the OPE, pole contribution and the masses of the $\bar{\Xi}_{bc}\Xi_{bc}$ and $\bar{\Xi}_{cc}\Xi_{bb}$ baryonium states are shown as functions of Borel parameter $M_B^2$ in Figs.~\ref{figbcbc0-+}$-$\ref{figbbcc1++}.

\begin{table}
\begin{center}
\renewcommand\tabcolsep{10pt}
\caption{The continuum thresholds, Borel parameters, and predicted masses of tetracharm and tetrabottom baryonium states}\label{mass_baryonium}
\begin{tabular}{ccccc}\hline\hline
Baryonium States                            &$J^{PC}$   & $\sqrt{s_0}\;(\text{GeV})$     &$M_B^2\;(\text{GeV}^2)$ &$M^X\;(\text{GeV})$       \\ \hline
 $c$-sector                                      &$0^{-+}$    & $7.6\pm0.2$                             &$5.3-6.3$                             &$7.33\pm0.12$         \\
                                                        &$1^{--}$     & $7.8\pm0.2$                             &$5.4-6.6$                             &$7.42\pm0.13$          \\
                                                        &$0^{++}$   &  $7.9\pm0.2$                             &$5.8-6.8$                             &$7.68\pm0.17$           \\
                                                        &$1^{++}$   &  $8.2\pm0.2$                             &$6.2-7.3$                             &$7.76\pm0.12$          \\\hline
 $b$-sector                                     &$0^{-+}$    & $19.9\pm0.2$                           &$14.2-18.1$                         &$19.41\pm0.15$        \\
                                                        &$1^{--}$     & $20.0\pm0.2$                            &$15.1-18.9$                         &$19.48\pm0.15$          \\
                                                       &$0^{++}$   &  $20.3\pm0.2$                           &$16.5-19.8$                         &$19.77\pm0.19$        \\
                                                       &$1^{++}$   &  $20.4\pm0.2$                             &$16.9-20.7$                           &$19.84\pm0.19$          \\\hline
$\bar{\Xi}_{bc}\Xi_{bc}$                 &$0^{-+}$    & $14.3\pm0.2$                        &$10.7-12.8$                         &$13.90\pm0.13$        \\
                                                      &$1^{--}$     & $14.5\pm0.2$                            &$12.4-14.8$                         &$13.96\pm0.15$          \\
                                                     &$0^{++}$   &  $14.6\pm0.2$                           &$11.5-13.6$                         &$14.29\pm0.10$        \\
                                                     &$1^{++}$   &  $14.9\pm0.2$                             &$14.1-16.2$                           &$14.36\pm0.14$          \\    \hline
$\bar{\Xi}_{cc}\Xi_{bb}$               &$0^{-+}$    & $14.0\pm0.2$                        &$10.7-12.5$                         &$13.53\pm0.14$        \\
                                                    &$1^{--}$     & $14.1\pm0.2$                            &$11.2-13.1$                         &$13.56\pm0.18$          \\
                                                    &$0^{++}$   &  $14.4\pm0.2$                           &$11.7-13.7$                         &$13.91\pm0.15$        \\
                                                   &$1^{++}$   &  $14.5\pm0.2$                             &$12.4-14.4$                           &$13.96\pm0.18$          \\
\hline
 \hline
\end{tabular}
\end{center}
\end{table}

\section{Decay analyses}\label{Decay}
To finally ascertain these tetra-heavy baryonium states, the straightforward procedure is to reconstruct them from its decay products, though the detailed characters still ask for more investigation. In our evaluation, the masses of tetracharm baryonium states are above the threshold of their respective $\bar{\Xi}_{cc}\Xi_{cc}$ dibaryons, so the $\bar{\Xi}_{cc}\Xi_{cc}$ decay channel will be the primary decay mode of tetracharm baryonium states. The typical decay modes of the tetra-heavy baryonium for different quantum numbers are given in Table \ref{decay-mode}, and these processes are expected to be measurable in the running LHC experiments.

It should be also be noted that $X(7200)$ is close in magnitude to the $0^{-+}$ tetracharm hybrid state calculated in Ref.~\cite{Wan:2020fsk}. In fact, the dynamical gluon of the hybrid can easily split into a quark pair. The main differences between the decay modes of tetracharm baryonium state and tetracharm hybris state is that the decay mode of di-$J/\psi$ for baryonium are suppressed by OZI, while the hybrid may decay to di-$J/\psi$ more straightforwardly.

\begin{table}
\begin{center}\caption{Typical decay modes of the tetra-heavy baryomium for each quantum number.}\label{decay-mode}
\renewcommand\tabcolsep{10pt}
\begin{tabular}{cccccccc}
\hline
\hline
                                            &$0^{-+}$                                     & $1^{--}$                                               &$0^{++}$                                &$1^{++}$   \\\hline
$c$-sector                           &$\bar{\Xi}_{cc}\Xi_{cc}$             &$\bar{\Xi}_{cc}\Xi_{cc}$                        &$\bar{\Xi}_{cc}\Xi_{cc}$         &$\bar{\Xi}_{cc}\Xi_{cc}$ \\
		                           &$J/\psi J/\psi \pi$                       &$J/\psi J/\psi \rho(770)$                         &$J/\psi J/\psi f_0(500)$          &$J/\psi J\psi \phi(1020)$  \\
		                           &$J/\psi J/\psi \eta$                     &$J/\psi J/\psi \omega(782)$                   &$J/\psi J/\psi f_0(980)$          &$J/\psi J/\psi f_1(1285)$  \\
		                          &$\eta_c \eta_c \pi$                     &$\eta_c \eta_c \rho(770)$                      &$\eta_c \eta_c f_0(500)$      &$\eta_c \eta_c \psi(1020)$ \\
		                          &$\eta_c \eta_c \eta$                   &$\eta_c \eta_c \omega(782)$                &$\eta_c \eta_c f_0(980)$      &$\eta_c \eta_c f_1(1285)$ \\
	                                  &$\eta_c D\bar{D}$                       &$J/\psi D\bar{D}$ 	                          &$\chi_{c0} D\bar{D}$              &$\chi_{c1} D\bar{D}$      \\\hline
$b$-sector                          &$\Upsilon \Upsilon \pi$               &$\Upsilon \Upsilon \rho(770)$               &$\Upsilon \Upsilon f_0(500)$      &$\Upsilon \Upsilon \phi(1020)$  \\
		                          &$\Upsilon \Upsilon\eta$              &$\Upsilon \Upsilon \omega(782)$         &$\Upsilon \Upsilon f_0(980)$      &  $\eta_b \eta_b \phi(1020)$\\
		                          &$\eta_b \eta_b \pi$                     &$\eta_b \eta_b \rho(770)$                      &$\eta_b \eta_b f_0(500)$            &$\chi_{b1} B\bar{B}$ \\
		                          &$\eta_b \eta_b \eta$                   &$\eta_b \eta_b \omega(782)$                &$\eta_b \eta_b f_0(980)$             & \\
	                                   &$\eta_b B\bar{B}$                      &$\Upsilon B\bar{B}$ 	                            &$\chi_{b0} B\bar{B}$                   &      \\\hline
$\bar{\Xi}_{bc}\Xi_{bc}$     &$B_c^{\ast +} B_c^{\ast -} \pi$    &$B_c^{\ast +} B_c^{\ast -} \rho(770)$    &$B_c^{\ast +} B_c^{\ast -} f_0(500)$      &$B_c^{\ast +} B_c^{\ast -} \phi(1020)$  \\
		                          &$B_c^{\ast +} B_c^{\ast -}\eta$   &$B_c^{\ast +} B_c^{\ast -} \omega(782)$ &$B_c^{\ast +} B_c^{\ast -} f_0(980)$      &  $B_c^{+} B_c^{-} \phi(1020)$\\
		                          &$B_c^+ B_c^- \pi$                     &$B_c^+ B_c^- \rho(770)$                        &$B_c^+ B_c^- f_0(500)$            & \\
		                          &$B_c^+ B_c^- \eta$                   &$B_c^+ B_c^- \omega(782)$                  &$B_c^+ B_c^- f_0(980)$             & \\
	                                   &$B_c^+ \bar{B}\bar{D}$              &$B_c^{\ast +} \bar{B}\bar{D}$ 	                           &                   &      \\
	                                   &$B_c^- BD$                                 &$B_c^{\ast -} BD$ 	                           &                   &      \\\hline
$\bar{\Xi}_{cc}\Xi_{bb}$     &$B_c^{\ast -} B_c^{\ast -} \pi$    &$B_c^{\ast -} B_c^{\ast -} \rho(770)$    &$B_c^{\ast -} B_c^{\ast -} f_0(500)$      &$B_c^{\ast -} B_c^{\ast -} \phi(1020)$  \\
		                          &$B_c^{\ast -} B_c^{\ast -}\eta$   &$B_c^{\ast -} B_c^{\ast -} \omega(782)$ &$B_c^{\ast -} B_c^{\ast -} f_0(980)$      &  $B_c^{-} B_c^{-} \phi(1020)$\\
		                          &$B_c^- B_c^- \pi$                     &$B_c^- B_c^- \rho(770)$                        &$B_c^- B_c^- f_0(500)$            & \\
		                          &$B_c^- B_c^- \eta$                   &$B_c^- B_c^- \omega(782)$                  &$B_c^- B_c^- f_0(980)$             & \\
	                                   &$B_c^- \bar{B}\bar{D}$             &$B_c^{\ast -} \bar{B}\bar{D}$ 	                           &                   &      \\\hline
\hline
\end{tabular}
\end{center}
\end{table}

\section{Conclusions}

In the framework of QCD sum rules we have investigated the tetracharm baryonium states in molecular configuration. Our numerical results indicate that there exist four possible tetracharm baryonium states with with masses $(7.33\pm0.12)$, $(7.42\pm0.13)$, $(7.68\pm0.17)$, and $(7.76\pm0.12)$ GeV for quantum numbers of $J^{PC}=0^{-+}$, $1^{--}$, $0^{++}$ and $1^{++}$,  respectively. Results indicate that the $X(7200)$ \cite{Aaij:2020fnh,ATLAS,CMS} is close in magnitude to our calculation of $0^{-+}$ tetracharm baryonium state. Moreover, the b-sector partners are also evaluated, and we find four tetra-bottom baryonium states exist with masses $(19.41\pm 0.15)$, $(19.48\pm 0.15)$, $(19.77\pm 0.19)$ and $(19.84\pm 0.19)$ GeV for $0^{-+}$, $1^{--}$, $0^{++}$ and $1^{++}$, respectively. Additionally, we also analyze the spectra of the $\bar{\Xi}_{bc}\Xi_{bc}$ and $\bar{\Xi}_{cc}\Xi_{bb}$ baryonium states, and their masses are tabulated in Table \ref{mass_baryonium}. Moreover, the primary and potential decay modes of these tetra-heavy baryonium states are analyzed, which might serve as a guide for experimental exploration in LHC.

Last, it should be noted that the pseudoscalar tetracharm baryonium was once investigated in the framework of QCD sum rule by Wang \cite{Wang:2021wjd} with different interpolating current and hence the result there has about $0.1$ GeV difference from this work.

\vspace{.5cm} {\bf Acknowledgments} \vspace{.5cm}

This work was supported in part by the National Key Research and Development Program of China under Contracts Nos. 2020YFA0406400, and the National Natural Science Foundation of China (NSFC) under the Grants 11975236 and 11635009. We are grateful to Z.G. Wang for finding a bug in the initial version.


\begin{widetext}
\appendix

\section{The spectral densities of double hidden-charm baryonium}

\subsection{The spectral densities for $0^{-+}$ and $0^{++}$ double hidden-charm baryonium}
For the current with the quantum number of $0^{-+}$ and $0^{++}$, we obtain the spectral
densities double hidden-charm baryonium as follows:
\begin{eqnarray}
\rho^{pert}(s)&=&\frac{1}{2^{16}\times 5\times 7 \pi^{10}} \int^{x_{+}}_{x_{-}} d x \int^{y_{+}}_{y_{-}} d y \int^{z_{+}}_{z_{-}} d z \int^{w_{+}}_{w_{-}} d w H_{xyzw}^5\nonumber\\
&\times&\Big[21A_{xyzw}^2 m_c^4  + xyzw H_{xyzw}^2 -7A_{xyzw}H_{xyzw}m_c^2(xy+wz) \Big]\;,\\
 \rho^{\langle \bar{q} q \rangle}(s) &=&0\;,\\
\rho^{\langle G^2 \rangle}(s)&=&\frac{\langle g_s^2 G^2 \rangle}{2^{18}\times \pi^{10}} \int^{x_{+}}_{x_{-}} d x \int^{y_{+}}_{y_{-}} d y \int^{z_{+}}_{z_{-}} d z \int^{w_{+}}_{w_{-}} d w H_{xyzw}^2\nonumber\\
&\times&\bigg[\frac{H_{xyzw}}{2\times 5 xyzw}\Big(20A_{xyzw}^2 m_c^4 (xy+wz)+xyzw(xy+wz)H_{xyzw}^2\nonumber\\
&-&10A_{xyzw}H_{xyzw}m_c^2(x^2y^2+xyzw+w^2z^2)\Big)+\frac{m_c^2}{3x^3y^3z^3w^3}\Big(xyzw C_{xyzw}H^2_{xyzw}\nonumber\\
&+&6A_{xyzw}^2(C_{xyzw}m_c^4+xyzwD_{xyzw}H_{xyzw}m_c^2) -A_{xyzw}H_{xyzw}\nonumber\\
&\times& (3xyzwE_{xyzw}H_{xyzw}+4C_{xyzw}m_c^2(xy+wz)) \Big)\bigg]\;,\\
\rho^{\langle \bar{q} G q \rangle}(s)&=&0\;,\\
 \rho^{\langle \bar{q} q \rangle^2}(s) &=&\frac{\mathcal{N}_i\langle \bar{q} q \rangle^2}{3\times2^{10}\times\pi^6} \int^{x_{+}}_{x_{-}} d x \int^{y_{+}}_{y_{-}} d y \int^{z_{+}}_{z_{-}} d z F_{xyz}^2\bigg[ 6m_c^4-2m_c^2(xy+z B_{xyz})\nonumber\\
 &\times&\Big(2B_{xyz}F_{xyz}-3s+2(x+y+z)F_{xyz}\Big)+3xyzB_{xyz}\Big(B^2_{xyz}F^2_{xyz}+2s^2\nonumber\\
 &-&4(x+y+z)sF_{xyz}+(x+y+z)^2F^2_{xyz}\nonumber\\
 &+&2B_{xyz}F_{xyz}\big(-2s+(x+y+z)F_{xyz}\big)\Big) \bigg]\;,\\
\rho^{\langle g_s^3 G^3 \rangle}(s)&=&\frac{\langle G^3 \rangle}{3\times2^{20}\pi^{10}}\int^{x_{+}}_{x_{-}} d x \int^{y_{+}}_{y_{-}} d y \int^{z_{+}}_{z_{-}} d z \int^{w_{+}}_{w_{-}} d w \frac{H_{xyzw}}{x^4y^4z^4w^4}\bigg[xyzwH_{xyzw}^2\nonumber\\
&\times&(8G_{xyzw}m_c^2+xyzwC_{xyzw}H_{xyzw})+A^2_{xyzw}(24G_{xyzw}m_c^6\nonumber\\
&+&36m_c^4 xyzw C_{xyzw}H_{xyzw})-4H_{xyzw}m_c^2A_{xyzw}\Big(6G_{xyzw}m_c^2(xy+wz)\nonumber\\
&+&xyzwH_{xyzw}\big(w^3z^3(x^3+y^3)(xy+6wz)+x^3y^3(6xy+wz)(w^3+z^3)\big)\Big)\bigg]\;,\\
\rho^{\langle \bar{q} q \rangle\langle \bar{q} G q \rangle}(s)&=&\frac{\mathcal{N}_i\langle \bar{q} q \rangle\langle \bar{q} G q \rangle}{2^9\times\pi^6} \int^{x_{+}}_{x_{-}} d x \int^{y_{+}}_{y_{-}} d y \int^{z_{+}}_{z_{-}} d z\bigg[ 4xyzB_{xyz}F_{xyz}^3 -3F_{xyz}^2\nonumber\\
&\times&\Big(6sxyzB_{xyz}+m_c^2(xy-xz-yz-z^2+z)\Big)-s\Big(m_c^4+s^2xyzB_{xyz}\nonumber\\
&+&s m_c^2(xy-xz-yz-z^2+z)\Big)+2F_{xyz}\Big(m_c^4+6s^2xyzB_{xyz}\nonumber\\
&+&3s m_c^2(xy-xz-yz-z^2+z)\Big)\bigg]\;,\\
 \rho^{\langle G^2 \rangle\langle \bar{q} q \rangle^2}(s) &=&\frac{\mathcal{N}_i\langle g_s^2 G^2 \rangle\langle \bar{q} q \rangle^2}{3\times2^{12}\pi^6} \int^{x_{+}}_{x_{-}} d x \int^{y_{+}}_{y_{-}} d y \int^{z_{+}}_{z_{-}} d z \frac{1}{xyzB_{xyz}}\bigg[2m_c^4(xy+zB_{xyz})\nonumber\\
& -&2m_c^2(x^2y^2+xyzB_{xyz}+z^2B_{xyz}^2)(2B_{xyz}F_{xyz}-s+2(x+y+z)F_{xyz})\nonumber\\
&+&xyzB_{xyz}(xy+zB_{xyz})\Big(3B_{xyz}^2F_{xyz}^2+s^2-6s(x+y+z)F_{xyz}\nonumber\\
&+&3(x+y+z)^2F_{xyz}^2+6B_{xyz}F_{xyz}(-s+(x+y+z)F_{xyz})\Big)\bigg]\;,
\end{eqnarray}
where $\mathcal{N}_i=-1$ and $+1$ for $0^{-+}$ and $0^{++}$ state, respectively. Here, we have the following definitions:
\begin{eqnarray}
&&A_{xyzw}=(1-x-y-z-w)\;,B_{xyz}=(1-x-y-z)\;,\\
&&H_{xyzw}=\bigg(\frac{1}{x}+\frac{1}{y}+\frac{1}{z}+\frac{1}{w}  \bigg)m_c^2-s\;,\\
&&C_{xyzw}=x^3y^3z^3+x^3y^3w^3+x^3z^3w^3+y^3z^3w^3\;,\\
&&D_{xyzw}=x^2y^2z^2+x^2y^2w^2+x^2z^2w^2+y^2z^2w^2\;,\\
&&E_{xyzw}=x^3y^3z^2+x^3y^3w^2+x^2z^3w^3+y^2z^3w^3\;,\\
&&F_{xyz}=\bigg(\frac{1}{x}+\frac{1}{y}+\frac{1}{z}+\frac{1}{1-x-y-z}  \bigg)m_c^2-s\;,\\
&&G_{xyzw}=x^4y^4z^4+x^4y^4w^4+x^4z^4w^4+y^4z^4w^4\;,\\
&&I_{xyzw}=x^4y^4z^3+x^4y^4w^3+x^3z^4w^4+y^3z^4w^4\;,\\
&&x_{\pm}=\bigg[\bigg( 1-\frac{8m_c^2}{s} \bigg) \pm \sqrt{\bigg( 1-\frac{8m_c^2}{s} \bigg)^2-\frac{4m_c^2}{s}}\bigg] \bigg/2\;,\\
&&y_{\pm}=\bigg[ 1+2 x +\frac{3 s x^2}{m_c^2-s x} \pm \sqrt{\frac{[m_c^2+s x (x-1)][(8x+1)m_c^2+s x(x-1)]}{(m_c^2-s x)^2}}  \bigg] \bigg/2\;,\\
&&z_{\pm}=\bigg[(1-x-y)\pm \sqrt{\frac{(x+y-1)[m_c^2(x+y-(x -y)^2)+s x y(x+y-1)]}{s x y-(x+y) m_c^2}} \bigg]\bigg/2\;,\\
&&w_{-}=\frac{x y z m_c^2}{s x y z -(x y +y z + x z)m_c^2}\;,w_{+}=1-x-y-z\;,
\end{eqnarray}
and all these definitions only can be used in Appendix A.

\subsection{The spectral densities for $1^{--}$ and $1^{++}$ double hidden-charm baryonium}
For the current with the quantum number of $1^{--}$ and $1^{++}$, we obtain the spectral
densities double hidden-charm baryonium as follows:
\begin{eqnarray}
\rho^{pert}(s)&=&\frac{1}{2^{17}\times 5\times 7 \pi^{10}} \int^{x_{+}}_{x_{-}} d x \int^{y_{+}}_{y_{-}} d y \int^{z_{+}}_{z_{-}} d z \int^{w_{+}}_{w_{-}} d w H_{xyzw}^5\nonumber\\
&\times&\Big[-14A_{xyzw}^3 m_c^4  + 2 xyzw H_{xyzw}^2 +7A_{xyzw}^2 m_c^2 (6m_c^2+xyH_{xyzw}\nonumber\\
&+&wzH_{xyzw}) -2A_{xyzw}H_{xyzw}\Big(xyzwH_{xyzw}+7m_c^2(xy+wz)\Big) \Big]\;,\\
\rho^{\langle \bar{q} q \rangle}(s) &=&0\;,\\
\rho^{\langle G^2 \rangle}(s)&=&\frac{\langle g_s^2 G^2 \rangle}{2^{19}\times \pi^{10}} \int^{x_{+}}_{x_{-}} d x \int^{y_{+}}_{y_{-}} d y \int^{z_{+}}_{z_{-}} d z \int^{w_{+}}_{w_{-}} d w H_{xyzw}^2\nonumber\\
&\times&\bigg[\frac{H_{xyzw}}{3\times 5 xyzw}\Big(-20A_{xyzw}^3 m_c^4 (xy+wz)+3xyzw(xy+wz)H^2_{xyzw}\nonumber\\
&-&3A_{xyzw}H_{xyzw}\big(xyzw(xy+wz)H_{xyzw}+ 10 m_c^2(x^2y^2+xyzw+w^2z^2)\big) \nonumber\\
&+& 15A_{xyzw}^2 \big(4m_c^4(xy+wz)+m_c^2H_{xyzw}(x^2y^2+xyzw+w^2z^2\big) \Big)\nonumber\\
&+&\frac{m_c^2}{2^3x^3y^3z^3w^3}\Big(2xyzw C_{xyzw}H^2_{xyzw}-4A_{xyzw}^3(xyzwm_c^2H_{xyzw}D_{xyzw}\nonumber\\
&+&m_c^4C_{xyzw})+A_{xyzw}^2(12C_{xyzw}m_c^4+3xyzw E_{xyzw}H^2_{xyzw}m_c^2 \nonumber\\
&+&4H_{xyzw}m_c^2(xyzwD_{xyzw}+(xy+wz)C_{xyzw}))-2 A_{xyzw}H_{xyzw}\nonumber\\
&\times&(xyzwH_{xyzw}(C_{xyzw}+3(xy+wz)D_{xyzw}+4C_{xyzw}m_c^2(xy+wz))) \Big)\bigg]\;,\\
\rho^{\langle \bar{q} G q \rangle}(s)&=&0\;,\\
 \rho^{\langle \bar{q} q \rangle^2}(s) &=&\frac{\mathcal{N}_i\langle \bar{q} q \rangle^2}{3\times2^{10}\times\pi^6} \int^{x_{+}}_{x_{-}} d x \int^{y_{+}}_{y_{-}} d y \int^{z_{+}}_{z_{-}} d z F_{xyz}^2\bigg[ 6m_c^4-2m_c^2(xy+z B_{xyz})\nonumber\\
 &\times&\Big(2F_{xyz}-3s\Big)+3B_{xyz}\Big(F_{xyz}^2-4sF_{xyz}+2s^2\Big) \bigg]\;,\\
\rho^{\langle g_s^3 G^3 \rangle}(s)&=&\frac{\langle G^3 \rangle}{3\times2^{20}\pi^{10}}\int^{x_{+}}_{x_{-}} d x \int^{y_{+}}_{y_{-}} d y \int^{z_{+}}_{z_{-}} d z \int^{w_{+}}_{w_{-}} d w \frac{H_{xyzw}}{x^4y^4z^4w^4}\bigg[xyzwH_{xyzw}^2\nonumber\\
&\times&(8G_{xyzw}m_c^2+xyzwC_{xyzw}H_{xyzw})-4A^3_{xyzw}(2G_{xyzw}m_c^6\nonumber\\
&+&3m_c^4 xyzw C_{xyzw}H_{xyzw})+2A_{xyzw}^2\Big(xyzwH^2_{xyzw}m_c^2\big((xy+wz)C_{xyzw}\nonumber\\
&+&5I_{xyzw}\big)+12m_c^6 G_{xyzw}+6m_c^4H_{xyzw}\big(3xyzwC_{xyzw}+(xy+wz)G_{xyzw}\big)\Big)\nonumber\\
&-&A_{xyzw}H_{xyzw}\Big(x^2y^2z^2w^2H_{xyzw}^2C_{xyzw}+4xyzw m_c^2 H_{xyzw}\big(6I_{xyzw}+2G_{xyzw}\nonumber\\
&+&xyzw(x^2y^2(w^3+z^3)+w^2z^2(x^3+y^3))\big)+24m_c^4(xy+wz)G_{xyzw}\Big)\bigg]\;,\\
\rho^{\langle \bar{q} q \rangle\langle \bar{q} G q \rangle}(s)&=&\mathcal{N}_i\frac{\langle \bar{q} q \rangle\langle \bar{q} G q \rangle}{2^9\times\pi^6} \int^{x_{+}}_{x_{-}} d x \int^{y_{+}}_{y_{-}} d y \int^{z_{+}}_{z_{-}} d z\bigg[ 4xyzB_{xyz}F_{xyz}^3 -3F_{xyz}^2\nonumber\\
&\times&\Big(6sxyzB_{xyz}+m_c^2(xy-xz-yz-z^2+z)\Big)-s\Big(m_c^4+s^2xyzB_{xyz}\nonumber\\
&+&s m_c^2(xy-xz-yz-z^2+z)\Big)+2F_{xyz}\Big(m_c^4+6s^2xyzB_{xyz}\nonumber\\
&+&3s m_c^2(xy-xz-yz-z^2+z)\Big)\bigg]\;,\\
 \rho^{\langle G^2 \rangle\langle \bar{q} q \rangle^2}(s) &=&\mathcal{N}_i\frac{\langle g_s^2 G^2 \rangle\langle \bar{q} q \rangle^2}{3\times2^{12}\pi^6} \int^{x_{+}}_{x_{-}} d x \int^{y_{+}}_{y_{-}} d y \int^{z_{+}}_{z_{-}} d z \frac{1}{xyzB_{xyz}}\bigg[2m_c^4(xy+zB_{xyz})\nonumber\\
& -&2m_c^2(x^2y^2+xyzB_{xyz}+z^2B_{xyz}^2)(2B_{xyz}F_{xyz}-s+2(x+y+z)F_{xyz})\nonumber\\
&+&xyzB_{xyz}(xy+zB_{xyz})\Big(3B_{xyz}^2F_{xyz}^2+s^2-6s(x+y+z)F_{xyz}\nonumber\\
&+&3(x+y+z)^2F_{xyz}^2+6B_{xyz}F_{xyz}(-s+(x+y+z)F_{xyz})\Big)\bigg]\;,
\end{eqnarray}
where $\mathcal{N}_i=-1$ and $+1$ for $1^{--}$ and $1^{++}$ state, respectively.

\section{The spectral densities of $\bar{\Xi}_{bc}\Xi_{bc}$ baryonium}

\subsection{The spectral densities for $0^{-+}$ and $0^{++}$ $\bar{\Xi}_{bc}\Xi_{bc}$ baryonium}
For the current with the quantum number of $0^{-+}$ and $0^{++}$, we obtain the spectral
densities of $\bar{\Xi}_{bc}\Xi_{bc}$ baryonium as follows:
\begin{eqnarray}
\rho^{pert}(s)&=&\frac{1}{2^{16}\times 5\times 7 \pi^{10}} \int^{x_{+}}_{x_{-}} d x \int^{y_{+}}_{y_{-}} d y \int^{z_{+}}_{z_{-}} d z \int^{w_{+}}_{w_{-}} d w H_{xyzw}^5\nonumber\\
&\times&\Big[21A_{xyzw}^2 m_c^2 m_b^2  + xyzw H_{xyzw}^2 -7A_{xyzw}H_{xyzw}m_c m_b (xy+wz) \Big]\;,\\
 \rho^{\langle \bar{q} q \rangle}(s) &=&0\;,\\
\rho^{\langle G^2 \rangle}(s)&=&\frac{\langle g_s^2 G^2 \rangle}{2^{18}\times \pi^{10}} \int^{x_{+}}_{x_{-}} d x \int^{y_{+}}_{y_{-}} d y \int^{z_{+}}_{z_{-}} d z \int^{w_{+}}_{w_{-}} d w H_{xyzw}^2\nonumber\\
&\times&\bigg[\frac{H_{xyzw}}{2\times 5 xyzw}\Big(20A_{xyzw}^2 m_c^2 m_b^2 (xy+wz)+xyzw(xy+wz)H_{xyzw}^2\nonumber\\
&-&10A_{xyzw}H_{xyzw}m_c m_b(x^2y^2+xyzw+w^2z^2)\Big)+\frac{1}{3x^3y^3z^3w^3}\Big(xyzw C_{xyzw}H_{xyzw}^2\nonumber\\
&+&6A_{xyzw}^2m_c^2m_b^2(C_{xyzw}+xyzwD_{xyzw}H_{xyzw}) -A_{xyzw}H_{xyzw}m_cm_b\nonumber\\
&\times& 4(xy+wz)C_{xyzw}-3H_n xyzw E_{xyzw}   \Big)\bigg]\;,\\
\rho^{\langle \bar{q} G q \rangle}(s)&=&0\;,\\
 \rho^{\langle \bar{q} q \rangle^2}(s) &=&\frac{\mathcal{N}_i\langle \bar{q} q \rangle^2}{3\times2^{10}\times\pi^6} \int^{x_{+}}_{x_{-}} d x \int^{y_{+}}_{y_{-}} d y \int^{z_{+}}_{z_{-}} d z F_{xyz}^2\bigg[ 6m_c^2m_b^2-2m_c m_b(xy+z B_{xyz})\nonumber\\
 &\times&\Big(2B_{xyz}F_{xyz}-3s+2(x+y+z)F_{xyz}\Big)+3xyzB_{xyz}\Big(B^2_{xyz}F^2_{xyz}+2s^2\nonumber\\
 &-&4(x+y+z)sF_{xyz}+(x+y+z)^2F^2_{xyz}\nonumber\\
 &+&2B_{xyz}F_{xyz}\big(-2s+(x+y+z)F_{xyz}\big)\Big) \bigg]\;,\\
\rho^{\langle g_s^3 G^3 \rangle}(s)&=&\frac{\langle G^3 \rangle}{3\times2^{20}\pi^{10}}\int^{x_{+}}_{x_{-}} d x \int^{y_{+}}_{y_{-}} d y \int^{z_{+}}_{z_{-}} d z \int^{w_{+}}_{w_{-}} d w \frac{H_{xyzw}}{x^4y^4z^4w^4}\bigg[xyzwH_{xyzw}^2\nonumber\\
&\times&(8G_{xyzw}+xyzwI_{xyzw}H_{xyzw})+12m_c^2m_b^2A^2_{xyzw}(2G_{xyzw}\nonumber\\
&+&3 xyzw I_{xyzw}H_{xyzw})-4H_{xyzw}m_cm_bA_{xyzw}\Big(6G_{xyzw}(xy+wz)\nonumber\\
&+&xyzwH_{xyzw}\big(w^3z^3(x^3+y^3)(xy+6wz)+x^3y^3(6xy+wz)(w^3+z^3)\big)\Big)\bigg]\;,\\
\rho^{\langle \bar{q} q \rangle\langle \bar{q} G q \rangle}(s)&=&\frac{\mathcal{N}_i\langle \bar{q} q \rangle\langle \bar{q} G q \rangle}{2^9\times\pi^6} \int^{x_{+}}_{x_{-}} d x \int^{y_{+}}_{y_{-}} d y \int^{z_{+}}_{z_{-}} d z\bigg[ 4xyzB_{xyz}F_{xyz}^3 -3F_{xyz}^2\nonumber\\
&\times&\Big(6sxyzB_{xyz}+m_c m_b(xy-xz-yz-z^2+z)\Big)-s\Big(m_c^2 m_b^2+s^2xyzB_{xyz}\nonumber\\
&+&s m_c m_b(xy-xz-yz-z^2+z)\Big)+2F_{xyz}\Big(m_c^2m_b^2+6s^2xyzB_{xyz}\nonumber\\
&+&3s m_c^2(xy-xz-yz-z^2+z)\Big)\bigg]\;,\\
 \rho^{\langle G^2 \rangle\langle \bar{q} q \rangle^2}(s) &=&\frac{\mathcal{N}_i\langle g_s^2 G^2 \rangle\langle \bar{q} q \rangle^2}{3\times2^{12}\pi^6} \int^{x_{+}}_{x_{-}} d x \int^{y_{+}}_{y_{-}} d y \int^{z_{+}}_{z_{-}} d z \frac{1}{xyzB_{xyz}}\bigg[2m_c^2m_b^2(xy+zB_{xyz})\nonumber\\
& -&2m_cm_b(x^2y^2+xyzB_{xyz}+z^2B_{xyz}^2)(2B_{xyz}F_{xyz}-s+2(x+y+z)F_{xyz})\nonumber\\
&+&xyzB_{xyz}(xy+zB_{xyz})\Big(3B_{xyz}^2F_{xyz}^2+s^2-6s(x+y+z)F_{xyz}\nonumber\\
&+&3(x+y+z)^2F_{xyz}^2+6B_{xyz}F_{xyz}(-s+(x+y+z)F_{xyz})\Big)\bigg]\;,
\end{eqnarray}
where $\mathcal{N}_i=-1$ and $+1$ for $0^{-+}$ and $0^{++}$ state, respectively. Here, we have the following definitions:
\begin{eqnarray}
&&A_{xyzw}=(1-x-y-z-w)\;,B_{xyz}=(1-x-y-z)\;,\\
&&H_{xyzw}=\bigg(\frac{1}{x}+\frac{1}{z}\bigg)m_c^2+\bigg(\frac{1}{y}+\frac{1}{w}  \bigg)m_b^2-s\;,\\
&&C_{xyzw}=m_c^2(w^3+x^3)y^3z^3+m_b^2(y^3+z^3)w^3x^3\;,\\
&&D_{xyzw}=x^2y^2z^2+x^2y^2w^2+x^2z^2w^2+y^2z^2w^2\;,\\
&&E_{xyzw}=x^3y^3z^2+x^3y^3w^2+x^2z^3w^3+y^2z^3w^3\;,\\
&&F_{xyz}=\bigg(\frac{1}{x}+\frac{1}{z}\bigg)m_c^2+\bigg(\frac{1}{y}+\frac{1}{1-x-y-z}  \bigg)m_b^2-s\;,\\
&&G_{xyzw}=m_c^2(x^4+w^4)y^4z^4+m_b^2(y^4+z^4)x^4w^4\;,\\
&&I_{xyzw}=x^3y^3z^3+x^3y^3w^3+x^3z^3w^3+y^3z^3w^3\;,\\
&&x_{\pm}=\bigg[\bigg( s-4m_b^2-4m_cm_b \bigg) \pm \sqrt{\bigg( s-4m_b^2-4m_cm_b \bigg)^2-4m_c^2s}\bigg] \bigg/2s\;,\\
&&y_{\pm}=\frac{1}{2(m_c^2-sx)}\bigg[m_c^2+2 x m_c m_b-sx+sx^2\nonumber\\
&&\pm\sqrt{4m_b^2(sx-m_c^2)(x^2-x)+(m_c^2+2 x m_c m_b-sx+sx^2)^2}\bigg]\;,\\
&&z_{\pm}=\frac{-1}{2\Big(x m_b^2+(m_c^2-sx)y\Big)}\bigg[-x m_b^2+x^2 m_b^2 -y m_c^2+2xym_c^2+sxy-sx^2y\nonumber\\
&&+y^2m_c^2-sxy^2 \pm\bigg(4xym_c^2(x+y-1)(xm_b^2+ym_c^2-sxy)+\Big(m_b^2(x-1)x\nonumber\\
&&+sxy(x+y-1)+ym_c^2(2x+y-1)\Big)^2 \bigg)^{1/2} \bigg]\;,\\
&&w_{-}=\frac{x y z m_b^2}{s x y z -(x y +y z )m_c^2- x z m_b^2}\;,w_{+}=1-x-y-z\;,
\end{eqnarray}
and all these definitions only can be used in Appendix B.

\subsection{The spectral densities for $1^{--}$ and $1^{++}$ $\bar{\Xi}_{bc}\Xi_{bc}$ baryonium}
For the current with the quantum number of $1^{--}$ and $1^{++}$, we obtain the spectral
densities as follows:
\begin{eqnarray}
\rho^{pert}(s)&=&\frac{1}{2^{17}\times 5\times 7 \pi^{10}} \int^{x_{+}}_{x_{-}} d x \int^{y_{+}}_{y_{-}} d y \int^{z_{+}}_{z_{-}} d z \int^{w_{+}}_{w_{-}} d w H_{xyzw}^5\nonumber\\
&\times&\Big[-14A_{xyzw}^3 m_c^2 m_b^2  + 2 xyzw H_{xyzw}^2 +7A_{xyzw}^2 m_c m_b (6m_c m_b+xyH_{xyzw}\nonumber\\
&+&wzH_{xyzw}) -2A_{xyzw}H_{xyzw}\Big(xyzwH_{xyzw}+7m_c m_b(xy+wz)\Big) \Big]\;,\\
\rho^{\langle \bar{q} q \rangle}(s) &=&0\;,\\
\rho^{\langle G^2 \rangle}(s)&=&\frac{\langle g_s^2 G^2 \rangle}{2^{19}\times \pi^{10}} \int^{x_{+}}_{x_{-}} d x \int^{y_{+}}_{y_{-}} d y \int^{z_{+}}_{z_{-}} d z \int^{w_{+}}_{w_{-}} d w H_{xyzw}^2\nonumber\\
&\times&\bigg[\frac{H_{xyzw}}{3\times 5 xyzw}\Big(-20A_{xyzw}^3 m_c^2 m_b^2 (xy+wz)+3xyzw(xy+wz)H^2_{xyzw}\nonumber\\
&-&3A_{xyzw}H_{xyzw}\big(xyzw(xy+wz)H_{xyzw}+ 10 m_c m_b(x^2y^2+xyzw+w^2z^2)\big) \nonumber\\
&+& 15A_{xyzw}^2 \big(4m_c^2 m_b^2(xy+wz)+m_c m_b H_{xyzw}(x^2y^2+xyzw+w^2z^2\big) \Big)\nonumber\\
&+&\frac{1}{2^3x^3y^3z^3w^3}\Big(2xyzw C_{xyzw}H^2_{xyzw}-4A_{xyzw}^3m_c^2 m_b^2(xyzwH_{xyzw}D_{xyzw}\nonumber\\
&+&C_{xyzw})+A_{xyzw}^2m_cm_b(12m_c m_b(C_{xyzw}+H_{xyzw}D_{xyzw})\nonumber\\
&+&3xyzw E_{xyzw}H^2_{xyzw}+4H_{xyzw}(xy+wz)C_{xyzw})-2 A_{xyzw}H_{xyzw}\nonumber\\
&\times&(xyzwH_{xyzw}(C_{xyzw}+3m_b m_c E_{xyzw})+4C_{xyzw}m_cm_b(xy+wz)) \Big)\bigg]\;,\\
\rho^{\langle \bar{q} G q \rangle}(s)&=&0\;,\\
 \rho^{\langle \bar{q} q \rangle^2}(s) &=&\frac{\mathcal{N}_i\langle \bar{q} q \rangle^2}{3\times2^{10}\times\pi^6} \int^{x_{+}}_{x_{-}} d x \int^{y_{+}}_{y_{-}} d y \int^{z_{+}}_{z_{-}} d z F_{xyz}^2\bigg[ 6m_c^2 m_b^2-2m_c m_b(xy+z B_{xyz})\nonumber\\
 &\times&\Big(2F_{xyz}-3s\Big)+3B_{xyz}\Big(F_{xyz}^2-4sF_{xyz}+2s^2\Big) \bigg]\;,\\
\rho^{\langle g_s^3 G^3 \rangle}(s)&=&\frac{\langle G^3 \rangle}{3\times2^{20}\pi^{10}}\int^{x_{+}}_{x_{-}} d x \int^{y_{+}}_{y_{-}} d y \int^{z_{+}}_{z_{-}} d z \int^{w_{+}}_{w_{-}} d w \frac{H_{xyzw}}{x^4y^4z^4w^4}\bigg[4A_{xyzw}m_c m_b^3 \nonumber\\
&\times&x^4 y^4 w^4\big( -2(A_{xyzw}-3)A_{xyzw}m_c m_b+3(A_{xyzw}-2)xyH_{xyzw} \big)\nonumber\\
&+&4m_b x^4 y^4 w^4 z H_{xyzw} \Big(  3A_{xyzw}m_c m_b \big((3-A_{xyzw})m_c A_{xyzw}+(A_{xyzw}-2)w m_b\big)\nonumber\\
 &+&H_{xyzw}\big( 3(A_{xyzw}-2) A_{xyzw}m_c +2(1-A_{xyzw})wm_b         \big)xy \Big)   +H_{xyzw}^2x^4 y^4 w^5 z^2\nonumber\\
 &\times& \Big((2A_{xyzw}-4)A_{xyzw}m_c m_b+(1-A_{xyzw})H_{xyzw}xy      \Big) -2A_{xyzw}m_c m_b z^4 \nonumber\\
 &\times& \Big((4A_{xyzw}-12)A_{xyzw}m_b^3 m_c x^4 w^4  +6H_{xyzw}m_b x^4 w^4 y\big( (A_{xyzw}-3)m_c A_{xyzw}\nonumber\\
 &+&(2-A_{xyzw}x m_b)     \big) -(A_{xyzw}-2)w^4x^5y^2H_{xyzw} +2(A_{xyzw}-3)A_{xyzw} y^4 m_c m_b\nonumber\\
 &\times& \big(3xwH_{xyzw}(w^3+x^3)+2m_c^2(w^4+x^4)\big) +  (2-A_{xyzw})xy^5H_{xyzw}\big(H_{xyzw}wx(w^3\nonumber\\
 &+&6x^3)+6m_c^2(w^4+x^4)\big)     \Big)  + wy^5H_{xyzw}\Big( (12A_{xyzw}-24)A_{xyzw}w^4x^4m_b^3m_c  \nonumber\\
 &+&4H_{xyzw}m_b w^4 x^4 y \big((3A_{xyzw}-6)A_{xyzw}m_c+(2-2A_{xyzw})xm_b\big) \nonumber\\
 &+&(1-A_{xyzw})H_{xyzw}^2w^4x^5y^2 +(2A_{xyzw}-4)y^4A_{xyzw}m_b m_c\big(H_{xyzw}wx(6w^3+x^3)\nonumber\\
 &+&6m_c^2(w^4+x^4)\big) +(1-A_{xyzw})xy^5H_{xyzw}\big( H_{xyzw}wx(w^3+x^3) \nonumber\\
 &+&8m_c^2(x^4+w^4)       \big)    \Big)      \bigg]\;,\\
\rho^{\langle \bar{q} q \rangle\langle \bar{q} G q \rangle}(s)&=&\mathcal{N}_i\frac{\langle \bar{q} q \rangle\langle \bar{q} G q \rangle}{2^9\times\pi^6} \int^{x_{+}}_{x_{-}} d x \int^{y_{+}}_{y_{-}} d y \int^{z_{+}}_{z_{-}} d z\bigg[ 4xyzB_{xyz}F_{xyz}^3 -3F_{xyz}^2\nonumber\\
&\times&\Big(6sxyzB_{xyz}+m_c m_b(xy-xz-yz-z^2+z)\Big)-s\Big(m_c^2 m_b^2+s^2xyzB_{xyz}\nonumber\\
&+&s m_c m_b(xy-xz-yz-z^2+z)\Big)+2F_{xyz}\Big(m_c^2 m_b^2+6 s^2xyzB_{xyz}\nonumber\\
&+&3s m_c m_b(xy-xz-yz-z^2+z)\Big)\bigg]\;,\\
 \rho^{\langle G^2 \rangle\langle \bar{q} q \rangle^2}(s) &=&\mathcal{N}_i\frac{\langle g_s^2 G^2 \rangle\langle \bar{q} q \rangle^2}{3\times2^{12}\pi^6} \int^{x_{+}}_{x_{-}} d x \int^{y_{+}}_{y_{-}} d y \int^{z_{+}}_{z_{-}} d z \frac{1}{xyzB_{xyz}}\bigg[2m_c^2 m_b^2(xy+zB_{xyz})\nonumber\\
& -&2m_c m_b(x^2y^2+xyzB_{xyz}+z^2B_{xyz}^2)(2B_{xyz}F_{xyz}-s+2(x+y+z)F_{xyz})\nonumber\\
&+&xyzB_{xyz}(xy+zB_{xyz})\Big(3B_{xyz}^2F_{xyz}^2+s^2-6s(x+y+z)F_{xyz}\nonumber\\
&+&3(x+y+z)^2F_{xyz}^2+6B_{xyz}F_{xyz}(-s+(x+y+z)F_{xyz})\Big)\bigg]\;,
\end{eqnarray}
where $\mathcal{N}_i=-1$ and $+1$ for $1^{--}$ and $1^{++}$ state, respectively.

\section{The spectral densities of $\bar{\Xi}_{cc}\Xi_{bb}$ baryonium}

\subsection{The spectral densities for $0^{-+}$ and $0^{++}$ $\bar{\Xi}_{cc}\Xi_{bb}$ baryonium}
For the current with the quantum number of $0^{-+}$ and $0^{++}$, we obtain the spectral
densities as follows:
\begin{eqnarray}
\rho^{pert}(s)&=&\frac{1}{2^{16}\times 5\times 7 \pi^{10}} \int^{x_{+}}_{x_{-}} d x \int^{y_{+}}_{y_{-}} d y \int^{z_{+}}_{z_{-}} d z \int^{w_{+}}_{w_{-}} d w H_{xyzw}^5\nonumber\\
&\times&\Big[21A_{xyzw}^2 m_c^2 m_b^2   + xyzw H_{xyzw}^2 -7A_{xyzw}H_{xyzw}(m_b^2xy+m_c^2wz) \Big]\;,\\
 \rho^{\langle \bar{q} q \rangle}(s) &=&0\;,\\
\rho^{\langle G^2 \rangle}(s)&=&\frac{\langle g_s^2 G^2 \rangle}{2^{18}\times \pi^{10}} \int^{x_{+}}_{x_{-}} d x \int^{y_{+}}_{y_{-}} d y \int^{z_{+}}_{z_{-}} d z \int^{w_{+}}_{w_{-}} d w H_{xyzw}^2\nonumber\\
&\times&\bigg[\frac{H_{xyzw}}{2\times 5 xyzw}\Big(20A_{xyzw}^2 m_c^2m_b^2 (xy+wz)+xyzw(xy+wz)H_{xyzw}^2\nonumber\\
&-&5A_{xyzw}H_{xyzw}(m_b^2xy(2xy+wz)+m_c^2wz(xy+wz))\Big)\nonumber\\
&+&\frac{1}{3x^3y^3z^3w^3}\Big(xyzw C_{xyzw}H^2_{xyzw}
+6A_{xyzw}^2m_c^2m_b^2(C_{xyzw}+xyzwD_{xyzw}H_{xyzw})\nonumber\\
&-&A_{xyzw}H_{xyzw} (3xyzwE_{xyzw}H_{xyzw}+4C_{xyzw}(m_b^2xy+m_c^2wz)) \Big)\bigg]\;,\\
\rho^{\langle \bar{q} G q \rangle}(s)&=&0\;,\\
 \rho^{\langle \bar{q} q \rangle^2}(s) &=&\frac{\mathcal{N}_i\langle \bar{q} q \rangle^2}{3\times2^{10}\times\pi^6} \int^{x_{+}}_{x_{-}} d x \int^{y_{+}}_{y_{-}} d y \int^{z_{+}}_{z_{-}} d z F_{xyz}^2\bigg[ 6m_c^2m_b^2-2m_b^2xy\nonumber\\
 &\times&\Big(2B_{xyz}F_{xyz}-3s+2(x+y+z)F_{xyz}\Big)+zB_{xyz}\Big(3xyB_{xyz}^2F_{xyz}^2\nonumber\\
 &+&m_c^2 \big(6s-4(x+y+z)F_{xyz}\big) +3xy\big( 2s^2-4sF_{xyz}(x+y+z)\nonumber\\
 &+&(x+y+z)^2F_{xyz}^2 \big)+2B_{xyz}F_{xyz}\big(-2m_c^2 +3xy(-2s\nonumber\\
 &+&(x+y+z)F_{xyz})  \big)         \Big) \bigg]\;,\\
\rho^{\langle g_s^3 G^3 \rangle}(s)&=&\frac{\langle G^3 \rangle}{3\times2^{20}\pi^{10}}\int^{x_{+}}_{x_{-}} d x \int^{y_{+}}_{y_{-}} d y \int^{z_{+}}_{z_{-}} d z \int^{w_{+}}_{w_{-}} d w \frac{H_{xyzw}}{x^4y^4z^4w^4}\bigg[xyzwH_{xyzw}^2\nonumber\\
&\times&(8G_{xyzw}+xyzwI_{xyzw}H_{xyzw})+12A_{xyzw}^2m_c^2m_b^2\Big(2G_{xyzw}\nonumber\\
&+&3xyzwI_{xyzw}H_{xyzw}\Big)
-4H_{xyzw}A_{xyzw}\Big( 6m_c^4(x^4+y^4)z^5w^5\nonumber\\
&+&6m_b^4(w^4+z^4)x^5y^5+6xyzwm_b^2m_c^2(x^3y^3w^4+x^3y^3z^4\nonumber\\
&+&w^3z^3x^4+w^3z^3y^4)+xyzwH_{xyzw}\big(xym_b^2(6x^3y^3z^3+6x^3y^3w^3\nonumber\\
&+&x^3z^3w^3+y^3z^3w^3)+wzm_c^2(6x^3z^3w^3+6y^3z^3w^3\nonumber\\
&+&x^3y^3z^3+x^3y^3w^3)\big)\Big)\bigg]\;,\\
\rho^{\langle \bar{q} q \rangle\langle \bar{q} G q \rangle}(s)&=&\frac{\mathcal{N}_i\langle \bar{q} q \rangle\langle \bar{q} G q \rangle}{2^9\times\pi^6} \int^{x_{+}}_{x_{-}} d x \int^{y_{+}}_{y_{-}} d y \int^{z_{+}}_{z_{-}} d z\bigg[ 4xyzB_{xyz}F_{xyz}^3 -3F_{xyz}^2\nonumber\\
&\times&\Big(6sxyzB_{xyz}+m_c^2zB_{xyz}+xym_b^2\Big)-s(m_c^2+sxy)(m_b^2+szB_{xyz})\nonumber\\
&+&2F_{xyz}\Big(m_b^2(m_c^2+3sxy)+3szB_{xyz}(m_c^2+2sxy)\Big)\bigg]\;,\\
 \rho^{\langle G^2 \rangle\langle \bar{q} q \rangle^2}(s) &=&\frac{\mathcal{N}_i\langle g_s^2 G^2 \rangle\langle \bar{q} q \rangle^2}{3\times2^{12}\pi^6} \int^{x_{+}}_{x_{-}} d x \int^{y_{+}}_{y_{-}} d y \int^{z_{+}}_{z_{-}} d z \frac{1}{xyzB_{xyz}}\bigg[2m_c^2 m_b^2(xy+zB_{xyz})\nonumber\\
& -&xym_b^2(2xy+zB_{xyz})(2B_{xyz}F_{xyz}-s+2(x+y+z)F_{xyz})\nonumber\\
&+&zB_{xyz}\Big( 3xyzB_{xyz}^3F_{xyz}^2 + xy\Big[  m_c^2(s-2(x+y+z)F_{xyz})+ xy\big(s^2\nonumber\\
&-&6s(x+y+z)F_{xyz}+s(x+y+z)^2F_{xyz}^2\big)  \Big] + B_{xyz}^2F_{xyz} \Big[ 3xyF_{xyz}\nonumber\\
&\times&\big( 2z(y+z)+x(y+2z)   \big) -2z(2m_c^2+3sxy) \Big] +B_{xyz}\Big[  sz(2m_c^2+sxy)\nonumber\\
&+&3xyF_{xyz}^2\big(2x(y+z)^2 +z(y+z)^2+x^2(2y+z) \big)- 2F_{xyz}\big(  3sxy(x+z)(y+z) \nonumber\\
&+& m_c^2(2z^2+2yz+xy+2zx) \big)  \Big] \Big)\bigg]\;,
\end{eqnarray}
where $\mathcal{N}_i=-1$ and $+1$ for $0^{-+}$ and $0^{++}$ state, respectively. Here, we have the following definitions:
\begin{eqnarray}
&&A_{xyzw}=(1-x-y-z-w)\;,B_{xyz}=(1-x-y-z)\;,\\
&&H_{xyzw}=\bigg(\frac{1}{x}+\frac{1}{y}\bigg)m_c^2+\bigg(\frac{1}{z}+\frac{1}{w}  \bigg)m_b^2-s\;,\\
&&C_{xyzw}=m_c^2(x^3+y^3)z^3w^3+m_b^2(w^3+z^3)y^3x^3\;,\\
&&D_{xyzw}=x^2y^2z^2+x^2y^2w^2+x^2z^2w^2+y^2z^2w^2\;,\\
&&E_{xyzw}=m_b^2 x^3y^3z^2+m_b^2x^3y^3w^2+m_c^2 x^2z^3w^3+m_c^2 y^2z^3w^3\;,\\
&&F_{xyz}=\bigg(\frac{1}{x}+\frac{1}{y}\bigg)m_c^2+\bigg(\frac{1}{z}+\frac{1}{1-x-y-z}  \bigg)m_b^2-s\;,\\
&&G_{xyzw}=m_c^2 x^4y^4z^4+m_c^2 x^4y^4w^4+m_b^2x^4z^4w^4+m_b^2y^4z^4w^4\;,\\
&&I_{xyzw}=x^3y^3z^3+x^3y^3w^3+x^3z^3w^3+y^3z^3w^3\;,\\
&&x_{\pm}=\bigg[\bigg( 1-\frac{4m_b(m_b+m_c)}{s} \bigg) \pm \sqrt{\bigg( 1-\frac{4m_b(m_b+m_c)}{s} \bigg)^2-\frac{4m_c^2}{s}}\bigg] \bigg/2\;,\\
&&y_{\pm}=\frac{1}{2(sx-m_c^2)}\bigg[-m_c^2-4m_b^2x+2m_c^2x+sx-sx^2\nonumber\\
&&\pm\sqrt{4(m_c^2-sx)(m_c^2x-m_c^2x^2)+(m_c^2+4m_b^2x-2m_c^2x-sx+sx^2)^2}\bigg]\;,\\
&&z_{\pm}=\bigg[(1-x-y)\pm \Big(\frac{1}{s x y-(x+y) m_c^2}\big[(x+y-1)m_c^2(x+y-\nonumber\\
&&(x +y)^2)+ x y\big(s(x+y-1)+4m_b^2\big)\big]\Big)^{1/2} \bigg]\bigg/2\;,\\
&&w_{-}=\frac{x y z m_b^2}{s x y z -(y z + x z)m_c^2-xym_b^2}\;,w_{+}=1-x-y-z\;,
\end{eqnarray}
and all these definitions only can be used in Appendix C.

\subsection{The spectral densities for $1^{--}$ and $1^{++}$ $\bar{\Xi}_{cc}\Xi_{bb}$ baryonium}
For the current with the quantum number of $1^{--}$ and $1^{++}$, we obtain the spectral
densities as follows:
\begin{eqnarray}
\rho^{pert}(s)&=&\frac{1}{2^{17}\times 5\times 7 \pi^{10}} \int^{x_{+}}_{x_{-}} d x \int^{y_{+}}_{y_{-}} d y \int^{z_{+}}_{z_{-}} d z \int^{w_{+}}_{w_{-}} d w H_{xyzw}^5\nonumber\\
&\times&\Big[-14A_{xyzw}^3 m_c^2 m_b^2  + 2 xyzw H_{xyzw}^2 +7A_{xyzw}^2 (m_b^2 (6m_c^2+xyH_{xyzw})\nonumber\\
&+&wzm_c^2H_{xyzw}) -2A_{xyzw}H_{xyzw}\Big(xyzwH_{xyzw}+7m_b^2xy+7m_c^2wz\Big) \Big]\;,\\
\rho^{\langle \bar{q} q \rangle}(s) &=&0\;,\\
\rho^{\langle G^2 \rangle}(s)&=&\frac{\langle g_s^2 G^2 \rangle}{2^{20}\times \pi^{10}} \int^{x_{+}}_{x_{-}} d x \int^{y_{+}}_{y_{-}} d y \int^{z_{+}}_{z_{-}} d z \int^{w_{+}}_{w_{-}} d w H_{xyzw}^2\nonumber\\
&\times&\bigg[\frac{H_{xyzw}}{3\times 5 xyzw}\Big(-40A_{xyzw}^3 m_c^2 m_b^2 (xy+wz)+6xyzw(xy+wz)H^2_{xyzw}\nonumber\\
&-&6A_{xyzw}H_{xyzw}\big(5m_b^2xy(2xy+wz)+wz(xyH_{xyzw}(xy+wz)\nonumber\\
&+&5m_c^2(xy+2wz))\big) + 15A_{xyzw}^2 \big(wzm_c^2H_{xyzw}(xy+2wz)\nonumber\\
&+&m_b^2(8m_c^2(xy+wz)+xy(2xy+wz)H_{xyzw})\big) \Big)\nonumber\\
&+&\frac{1}{2^2x^3y^3z^3w^3}\Big(2xyzw C_{xyzw}H^2_{xyzw}-4A_{xyzw}^3m_c^2 m_b^2(xyzwH_{xyzw}D_{xyzw}\nonumber\\
&+&C_{xyzw})+A_{xyzw}^2(12C_{xyzw}m_c^2 m_b^2+3xyzw E_{xyzw}H^2_{xyzw} \nonumber\\
&+&4H_{xyzw}(3xyzwD_{xyzw}m_b^2m_c^2+(xym_b^2+wzm_c^2)C_{xyzw}))\nonumber\\
&-&2 A_{xyzw}H_{xyzw}\big(  4m_c^4w^4z^4(x^3+y^3)+4m_b^4x^4y^4(w^3+z^3)+4xyzwm_c^2m_b^2\nonumber\\
&\times&(w^3x^2y^2+w^2x^3z^2+w^2y^3z^2+x^2y^2z^3)\nonumber\\
& +&xyzwH_{xyzw}(3E_{xyzw}+C_{xyzw})
\big) \Big)\bigg]\;,\\
\rho^{\langle \bar{q} G q \rangle}(s)&=&0\;,\\
 \rho^{\langle \bar{q} q \rangle^2}(s) &=&\frac{\mathcal{N}_i\langle \bar{q} q \rangle^2}{3\times2^{10}\times\pi^6} \int^{x_{+}}_{x_{-}} d x \int^{y_{+}}_{y_{-}} d y \int^{z_{+}}_{z_{-}} d z F_{xyz}^2\bigg[ m_b^2(m_c^2-4xyF_{xyz}+6sxy)\nonumber\\
 &+&zB_{xyz}\Big(3xyF_{xyz}^2+6s(m_c^2+sxy)-4F_n(m_c^2+3sxy)\Big) \bigg]\;,\\
\rho^{\langle g_s^3 G^3 \rangle}(s)&=&\frac{\langle G^3 \rangle}{3\times2^{20}\pi^{10}}\int^{x_{+}}_{x_{-}} d x \int^{y_{+}}_{y_{-}} d y \int^{z_{+}}_{z_{-}} d z \int^{w_{+}}_{w_{-}} d w \frac{H_{xyzw}}{x^4y^4z^4w^4}\bigg[4A_{xyzw}m_b^4x^4y^4w^4\nonumber\\
&\times&(2(3-A_{xyzw}) A_{xyzw}m_c^2+3xy(A_{xyzw}-2)H_{xyzw})-4m_b^2H_{xyzw}x^4y^4w^4z\nonumber\\
&\times&\Big(3A_{xyzw}m_c^2\big(A_{xyzw}^2+2w-(w+3)A_{xyzw}\big)+xyH_{xyzw}\big(3(2-A_{xyzw})A_{xyzw}\nonumber\\
&+&2w(A_{xyzw}-1)\big)\Big)+H_{xyzw}^2w^5x^4y^4z^2(2(A_{xyzw}-2)A_{xyzw}m_c^2+(1\nonumber\\
&-&A_{xyzw}xyH_{xyzw}))-2m_b^2z^4A_{xyzw}\Big(4(A_{xyzw}-3)A_{xyzw}m_c^4w^4x^4+6H_{xyzw}\nonumber\\
&\times&m_c^2w^4x^4y(A_{xyzw}^2+2x-(x+3)A_{xyzw})+(A_{xyzw}-2)w^4x^5y^2H_{xyzw}^2\nonumber\\
&+&2(A_{xyzw}-3)A_{xyzw}m_c^2y^4(2m_c^2w^4+2m_b^2x^4+3H_{xyzw}wx(x^3+w^3))\nonumber\\
&+&(2-A_{xyzw})xy^5H_{xyzw}(6m_c^2w^4+H_{xyzw}w^4x+6m_b^2x^4+6H_{xyzw}wx^4)\Big)\nonumber\\
&+&wz^5H_{xyzw}\Big(12(A_{xyzw}-2)A_{xyzw}m_c^4w^4x^4+4H_{xyzw}m_c^2w^4x^4y(3A_{xyzw}\nonumber\\
&+&2x-2(x+3)A_{xyzw})+(1-A_{xyzw})H_{xyzw}^2w^4x^5y^2+2(A_{xyzw}-2)\nonumber\\
&\times&A_{xyzw}m_c^2y^4(6m_c^2w^4+6h_{xyzw}xw^4+6m_b^2x^4+H_{xyzw}wx^4)\nonumber\\
&+&(1-A_{xyzw})xy^5H_{xyzw}(8m_c^2w^4+8m_b^2x^4+H_{xyzw}wx(w^3+x^3))
\Big)
\bigg]\;,\\
\rho^{\langle \bar{q} q \rangle\langle \bar{q} G q \rangle}(s)&=&\mathcal{N}_i\frac{\langle \bar{q} q \rangle\langle \bar{q} G q \rangle}{2^9\times\pi^6} \int^{x_{+}}_{x_{-}} d x \int^{y_{+}}_{y_{-}} d y \int^{z_{+}}_{z_{-}} d z\bigg[ 4xyzB_{xyz}F_{xyz}^3 -3F_{xyz}^2\nonumber\\
&\times&\Big(xym_b^2+zB_{xyz}(m_c^2+6sxy)\Big)-s(m_c^2+sxy)(m_b^2+zsB_{xyz})\nonumber\\
&+&2F_{xyz}\Big(m_b^2(m_c^2+3sxy)+3szB_{xyz}(m_c^2+2sxy)\Big)\bigg]\;,\\
 \rho^{\langle G^2 \rangle\langle \bar{q} q \rangle^2}(s) &=&\mathcal{N}_i\frac{\langle g_s^2 G^2 \rangle\langle \bar{q} q \rangle^2}{3\times2^{12}\pi^6} \int^{x_{+}}_{x_{-}} d x \int^{y_{+}}_{y_{-}} d y \int^{z_{+}}_{z_{-}} d z \frac{1}{xyzB_{xyz}}\bigg[m_b^2\Big(2m_c^2(xy+zB_{xyz})\nonumber\\
 &-&xy(2xy+zB_{xyz})(2B_{xyz}F_{xyz}-s+2(x+y+z)F_{xyz})\Big)+zB_{xyz}\nonumber\\
 &\times&\Big(3xyzB_{xyz}^3F_{xyz}^2+xy\big(m_c^2(s-2(x+y+z)F_{xyz})+xy(s^2-6s(x+y\nonumber\\
 &+&z)F_{xyz}+3(x+y+z)^2F_{xyz}^2)\big)+B_{xyz}^2F_{xyz}\big(3xyF_{xyz}(2z^2+2yz+xy+2xz)\nonumber\\
 &-&2(2m_c^2+3sxy)\big)+B_{xyz}\big(sz(2m_c^2+sxy)+3xyF_{xyz}(2x(y+z)^2+z(y+z)^2\nonumber\\
 &+&(2y+z)x^2)-2F_{xyz}(3sxy(x+z)(y+z)\nonumber\\
 &+&m_c^2(2z(y+z)+x(y+z))) \big) \Big) \bigg]\;,
\end{eqnarray}
where $\mathcal{N}_i=-1$ and $+1$ for $1^{--}$ and $1^{++}$ state, respectively.

\end{widetext}
\end{document}